\pdfoutput=1

\documentclass[a4paper,fleqn,usenatbib]{mnras}
\usepackage[T1]{fontenc}
\usepackage{ae,aecompl}
\usepackage{pdflscape}
\usepackage{booktabs}
\usepackage{threeparttable}
\usepackage{supertabular}
\usepackage{graphicx}	
\usepackage{amsmath}	
\usepackage{amssymb}	
\usepackage{multirow}
\usepackage{framed}
\usepackage{ulem}
\usepackage{xcolor}
\usepackage{float}

\newcommand\redsout{\bgroup\markoverwith{\textcolor{red}{\rule[0.5ex]{2pt}{1pt}}}\ULon}
\def\rads{\hbox{\rm\hskip.35em rad s}$^{-1}$}
\def\qvorticity{\hbox{\rm\hskip.35em  cm$^{2}$ s}$^{-1}$}
\def\gcc{\hbox{\rm\hskip.35em  g cm}$^{-3}$}

\title[Pulse Profile and Glitch of PSR J0738$-$4042]{New pulse profile variability associated with a glitch of PSR J0738$-$4042}
\author[S. Q. Zhou et al.]
       {S. Q. Zhou,$^{1,4,8,9}$
        E. G\"{u}gercino\u{g}lu,$^{2,3}$\thanks{E-mail: \href{egugercinoglu@gmail.com}{egugercinoglu@gmail.com}}
        J. P. Yuan,$^{4,10}$\thanks{E-mail: \href{yuanjp@xao.ac.cn}{yuanjp@xao.ac.cn}}
        M. Y. Ge,$^{5}$
        C. Yu,$^{1,8,9}$\newauthor
        C. M. Zhang,$^{2,11,12}$
        J. Zhang,$^{6}$
        Z. W. Feng$^{7}$ and
        C. Q. Ye$^{1,9}$
        \\
$^{1}$School of Physics and Astronomy, Sun Yat-Sen University, Zhuhai, 519082, China\\
$^{2}$National Astronomical Observatories, Chinese Academy of Sciences, 20A Datun Road, Chaoyang District, Beijing 100101, China\\
$^{3}$Istanbul University, Faculty of Science, Department of Astronomy and Space Sciences, Beyaz{\i}t, 34119, Istanbul, Turkey\\
$^{4}$Xinjiang Astronomical Observatory, Chinese Academy of Sciences, Xinjiang 830011, China\\
$^{5}$Key Laboratory of Particle Astrophysics, Institute of High Energy Physics, Chinese Academy of Sciences, Beijing 100049, China\\
$^{6}$Department of Physics and Electronic Engineering, QiLu Normal University, Jinan 250033, China\\
$^{7}$School of Physics and Astronomy, China West Normal University, Nanchong, 637009, China\\
$^{8}$State Key Laboratory of Lunar and Planetary Sciences, Macau University of Science and Technology, Macau, China\\
$^{9}$CSST Science Centre for the Guangdong-Hong Kong-Macau Greater Bay Area, Zhuhai, 519082, China\\
$^{10}$Centre for Astronomical Mega-Science, Chinese Academy of Sciences, Beijing 100012, China\\
$^{11}$University of Chinese Academy of Sciences, School of Astronomy and Space Science, Beijing 100049, China\\
$^{12}$University of Chinese Academy of Sciences, School of Physical Sciences, Beijing 100049, China\\}

\date{Accepted XXX. Received YYY; in original form ZZZ}
\pubyear{2022}

\begin{document}  
\label{firstpage}
\pagerange{\pageref{firstpage}--\pageref{lastpage}}
\maketitle

\begin{abstract}
The close correlation observed between emission state and spin-down rate change of pulsars has many implications both for the magnetospheric physics and the neutron star interior. The middle-aged pulsar PSR J0738$-$4042, which had been observed to display variations in the pulse profile associated with its spin-down rate change due to external effects, is a remarkable example. In this study, based on the 12.5-yr combined public timing data from UTMOST and Parkes, we have detected a new emission-rotation correlation in PSR J0738$-$4042 concurrent with a glitch. A glitch that occurred at MJD 57359(5) (December 3, 2015) with $\Delta\nu/\nu \sim 0.36(4)\times 10^{-9}$ is the first glitch event observed in this pulsar and is probably the underlying cause of the emission-rotation correlation. Unlike the usual post-glitch behaviours, the braking torque on the pulsar has continued to increase over 1380 d, corresponding to a significant decrease in $\ddot{\nu}$. As for changes in the pulse profile after the glitch, the relative amplitude of the leading component weakens drastically, while the middle component becomes stronger. A combined model of crustquake induced platelet movement and vortex creep response is invoked to account for this rare correlation. In this scenario, magnetospheric state-change is naturally linked to the pulsar-intrinsic processes that give rise to a glitch.  
\end{abstract}




\section{Introduction}
Pulsars are highly-magnetized, rotating neutron stars. Following the discovery of pulsars, the technique known as ``pulsar timing'' has been naturally used to examine how they rotate. The long-term timing observations have uncovered two types of intrinsic irregularities in the pulsar rotational evolution: timing noise and glitches. Timing noise in most pulsars is significantly dominated by sustained random wandering in either the phase, spin, or spin-down rate \citep{bgh72,hlk10,ymw17}. Glitches are observed as abrupt increases in the rotation and the spin-down rates of pulsars instantaneous to the accuracy of the data \citep{pdh18,alg19}. In general, the post-glitch behaviour is modelled as an initial slow exponential recovery over a few days to several months followed by linear decay of the part of the increase in the spin-down rate on a time-scale of years \citep{sl96,ywm10,espinoza11,yu13,liu21,bsa22,egy22}. Pulsar glitches and timing noise can be used to probe into the neutron star internal structure and dynamics \citep{cheng87,jones90,ho15,sourie20,erbil20,montoli20}.  

Apart from the above mentioned rotational instabilities cyclic behaviour and switching between two or more states have been reported in increasing number of sources, and in some of them a correlation exists with pulse profile variation. \cite{lhk10} demonstrated that for six pulsars the spin-down rate variations are correlated with the pulse shape changes. Periodic modulation in the arrival times of some pulsars is evaluated as being due to precession \citep{kerr16}. The first glitch-triggered pulse profile changes were observed in the very young pulsar PSR J1119$-$6127 \citep{wje11}. Recently, \citet{shaw22} conducted a search for emission-correlated $\dot{\nu}$ transitions in 17 pulsars previously studied by \cite{lhk10} and identified a new correlation in PSR B1642$-$03. At present, radiative changes accompanied by timing irregularities are extremely rare in pulsars, except for the well-established case of magnetars \citep{dk14, man18}. This may be, in part due to the comparatively low quality of the available data on many pulsars (low observational cadence, low S/N), resulting in inability to detect the correlated slight variations in their rotational and emission properties.


\begin{table*}
\label{tab:Emission}
\caption{Detailed parameters for emission-rotation correlations in thirteen pulsars.}
\begin{center}
\begin{threeparttable}
\begin{tabular}{lllccccrrcc}
  
    \hline
    
\multicolumn{1}{c}{Pulsar Name}                  & \multicolumn{1}{c}{$P$}               & \multicolumn{1}{c}{$\tau_{c}$}                   & \multicolumn{1}{c}{$B_{s}$}           & \multicolumn{1}{c}{$\dot{E}$} 			         & \multicolumn{1}{c}{Glitch?}           & \multicolumn{1}{c}{$\Delta\nu/\nu$}              & \multicolumn{1}{c}{$\Delta\dot{\nu}/\dot{\nu}$} & \multicolumn{1}{c}{Flux?}                        & \multicolumn{1}{c}{Profile}           &
\multicolumn{1}{c}{Ref.}                         \\

\multicolumn{1}{c}{(PSR)}                        & \multicolumn{1}{c}{(s)}               & \multicolumn{1}{c}{(kyr)}                        & \multicolumn{1}{c}{($10^{12} \rm\ G $)} & \multicolumn{1}{c}{($10^{32}\rm \ erg/s$)}       & \multicolumn{1}{c}{(Y/N)}             & \multicolumn{1}{c}{($10^{-9}$)}                  & \multicolumn{1}{c}{($10^{-3}$)}       & 
\multicolumn{1}{c}{}                             & \multicolumn{1}{c}{}                  &
\multicolumn{1}{c}{}                             \\
        
    \hline
    
    J0738$-$4042      & 0.3749                & 4320                   & 0.727                & 10
& N                  & --                    & --140(--)              & --                   & $W_{50}$($\uparrow$)                         & 1 \\
    
                     &                       &                        &                      & 
& Y                  & 0.36(4)               & 3(1)                   & --                   & $W_{50}$($\downarrow$)                       & This work \\

    J0742$-$2822      & 0.1667                & 157                    & 1.69                 & 1400
& N                  & --                    & 6.6(--)                & --                   &  
$W_{\rm 75}$($\downarrow$)                   & 2 \\

                     &                       &                        &                      & 
& Y                  & 102.73(11)            & 2.1(5)                 & --                   & 
$\star$                                      & 3 - 4  \\

    J1001$-$5507      & 1.4366                & 441                    & 8.71                 & 6.9
& N                  & --                    & 13.0(3)                & --                   & 
$W_{\rm eq}$($\downarrow$)                   & 5     \\

    J1119$-$6127      & 0.4079                & 1.61                   & 41                   & 23000
& Y                  & 9400(300)             & 580(14)                & --                   &
$*$                                          & 6      \\

    J1543$-$0620      & 0.7090                & 12800                  & 0.799                & 0.97
& N                  & --                    & 17.1(--)               & --                   & 
$W_{\rm 10}$($\downarrow$)                   & 2 \\

    J1602$-$5100      & 0.8642                & 197                    & 7.84                 & 42
& N                  & --                    & 38.6(4)                & --                   & 
$W_{\rm 10}$($\uparrow$)                     & 7 - 8    \\

    B1822$-$09        & 0.7690                & 233                    & 6.42                 & 45
& N                  & --                    & 24.2(4)                & --                   & 
$A_{\rm pc}/A_{\rm mp}$($\uparrow$)          & 2, 9   \\

                     &                       &                        &                      & 
& N                  & --                    & 19(1)                  & --                   & 
$A_{\rm pc}/A_{\rm mp}$($\uparrow$)          & 2, 10   \\

                     &                       &                        &                      & 
&Y                   & 4.08(2)               & 0.08(1)                & --           
&{{$\diamond$}}               & {{11}}   \\

                     &                       &                        &                      & 
&Y                   & 7.2(1)                & 1.65(7)                & --           
&{{$\diamond$}}               & {{11}}   \\

    B1828$-$11        & 0.4050                & 107                    & 4.99                 & 360
& N                  & --                    & 7.1(--)                & --                   & 
$W_{\rm 10}$($\downarrow$)                   & 2    \\

    J2021+4026       & 0.2653                & 76.9                   & 3.85                 & 1200
& Y                  & <100(--)$^a$          & 56(9)                  & $\gamma$($\downarrow$)& 
Y                                            & 12 - 15 \\

                     &                       &                        &                      & 
& N                  & --                    & 31(11)                 & $\gamma$($\downarrow$)& 
Y                                            & 15 \\

    B2021+51         & 0.5291                & 2740                   & 1.29                 & 8.2
& Y                  & 0.373(5)              & --0.24(3)              & --                   & 
$W_{\rm 10}$($\downarrow$)                   & 16 \\

    B2035+36         & 0.6187                & 2180                   & 1.69                 & 7.5
& N                  & --                    & 132.8(--)              & --                   & 
$W_{\rm eq}$($\downarrow$)                   & 2   \\

                     &                       &                        &                      &
& Y                  & 7.7(8)                & 67(8)                  & --                   & 
$W_{50}$($\downarrow$)                       & 17   \\

   J2043+2740        & 0.0961                & 1200                   & 0.354                & 560
& N                  & --                    & 59.1(--)               & --                   & 
$W_{50}$($\uparrow$)                         & 2   \\

    \hline
    
\end{tabular}
\begin{tablenotes}
\item[] 
$^a$: denotes that a glitch event, its size, $\Delta\nu/\nu$, is less than $10^{-7}$. \\
$^b$: $A_{\rm pc}/A_{\rm mp}$ is the ratio of the amplitudes of the precursor and main pulse. \\
$^c$: $W_{\rm eq}$ is the pulse equivalent width (the ratio of the area under the pulse to the peak pulse amplitude).\\
$^d$: $W_{\rm 10}$, $W_{\rm 50}$ and $W_{\rm 75}$ are the full widths of the pulse profile at $10\%$, $50\%$, and $75\%$ of the peak pulse amplitude, respectively.\\
$\star$: represents that the appearance of additional pulse components is closely correlated with an unusual glitch.\\
$*$: indicates the correlation between the ratio of the two-components in the profile and $\Delta\dot{\nu}$, which rapidly increases after the glitch.\\
$\diamond$: implies the variations of the integrated mean pulse profiles in both the radio-bright (B-mode) and the radio-quiet (Q-mode) modes.\\
References for these parameters:  
1 -- \cite{bkb14}; 2 -- \cite{lhk10}; 3 -- \cite{ksj13}; 4 -- \cite{dww21}; 5 -- \cite{cb12}; 6 -- \cite{wje11}; 7 -- \cite{bkj16}; 8 -- \cite{zzz19}; 9 -- \cite{s07}; 10 -- \cite{zww04}; 11 -- \cite{lws22}; 12 -- \cite{abb13}; 13 -- \cite{ntc16}; 14 -- \cite{znl17}; 15 -- \cite{twl20}; 16 -- \cite{lwy21}; 17 -- \cite{kyw18}.\\

\end{tablenotes}
\end{threeparttable}
\end{center}
\end{table*}

Table \ref{tab:Emission} contains the detailed parameters for each observed emission-rotation correlation in thirteen normal pulsars known to us. Their emission mode-changing features -- sudden changes to the pulse profile shape or the total flux density, are associated with irregularities in the spin properties \citep{bkj16}. During this period, the spin-down rate usually jumps up to a higher state \citep{twl20,shaw22}. The observed sizes of glitches coupled with
emission mode switching decrease with increasing characteristic age $\tau_{c}$. Notably, PSR J0738$-$4042 is the first pulsar that showed a significant reduction in the spin-down rate interrelated with the emergence of a new profile component, hypothesised to be caused by an asteroid encounter \citep{krj11, bkb14}. A large glitch with two exponential recoveries in PSR J1119$-$6227 was found to be coincident with the appearance of additional pulse components with intermittent or RRAT-like behaviour \citep{wje11}. Repeated state changes have been detected in the only known variable gamma-ray pulsar PSR J2021+4026 \citep{twl20}. The gamma-ray pulsar PSR J1124$-$5916 has also experienced spin-down rate transition with no significant
difference between the pulse profiles in different states \citep{ge20a}. The pulse profile switching between wide and narrow modes is induced by a glitch occurred in PSR B2035+36 \citep{kyw18}. Overall, these complicated strong connections between emission and rotational properties can provide new insights into the magnetospheric conditions and neutron-star interiors.

The theoretical picture of these state-switching correlations remains ambiguous. Nevertheless, the scenario that the measurable changes in spin-down rate, flux, and pulse shape of pulsars is driven by a shift in the magnetic inclination angle $\alpha$ as a consequence of a glitch, has gradually become a consensus \citep{le97, akbal15, ntc16, lwy21}. In this work, motivated by the interesting repeated mode-changing behaviours in several pulsars \citep{kyw18, twl20} and a potential for periodical variations in the emission profile and $\dot{\nu}$ of PSR J0738$-$4042 \citep{bkb14}, we carry out data analysis of PSR J0738$-$4042 by using the combined UTMOST and Parkes timing observations conducted between March 2008 and September 2020 to keep track of the long-term rotational history and pulse profile evolution for this pulsar. Along the way a new glitch is identified associated with the pulse profile changes in PSR J0738$-$4042. The paper is organised as follows. Pulsar observations and data reduction methodology are outlined in Section \ref{observations}. Detailed results of glitch analysis and glitch-triggered emission variations are presented in Section \ref{glitch} and \ref{correlation}, respectively. In Section \ref{model}, the combined model of crustquake induced platelet movement and vortex creep response is applied to the observed data which enables to better explain the observed pulsar state-changing behaviour. Further discussions are made regarding these findings and their comparison with similar cases in the literature in Section \ref{discussion}. Finally in Section \ref{conclusions} the paper is closed with concluding remarks.

\section{Observations and Analysis}{\label{observations}}

PSR J0738$-$4042 is an isolated pulsar that was discovered 54 years ago in 1968 \citep{lvw68}, and has a spin period of $P = 374 \ \rm ms$ and a period derivative of $\dot{P} = 1.37\times10^{-15}$ \citep{manchester05,lbs20}. Considering the magnetic dipole radiation as the main cause of the spin-down and if the initial spin period is much less than its current value, the spin parameters imply that this pulsar has a large characteristic age $\tau_c =P/2\dot{P}= 4320 \rm \ kyr$, making it one of the oldest state-changing pulsars, with the surface magnetic field $B_s$ and spin-down energy loss rate $\dot{E}$ of $0.72\times10^{12}\ \rm G$ and $10\times10^{32}\ \rm erg/s$, respectively\footnote{\href{https://www.atnf.csiro.au/research/pulsar/psrcat/}{https://www.atnf.csiro.au/research/pulsar/psrcat/}} \citep{manchester05,lbs20}. The mean timing noise parameter $\Delta_8$ [$\equiv\log(|\ddot{\nu}|t^3/6\nu)$, \citet{arzoumanian94}] for PSR J0738$-$4042 is calculated as $-1.14(2)$  with our data, and when compared with those of 366 pulsars in \citet{hlk10} indicates a high level of timing noise. This pulsar had never been reported to experience a glitch before \citep{bkj16}.

\begin{figure*}
\centering
\includegraphics[width=0.96\textwidth]{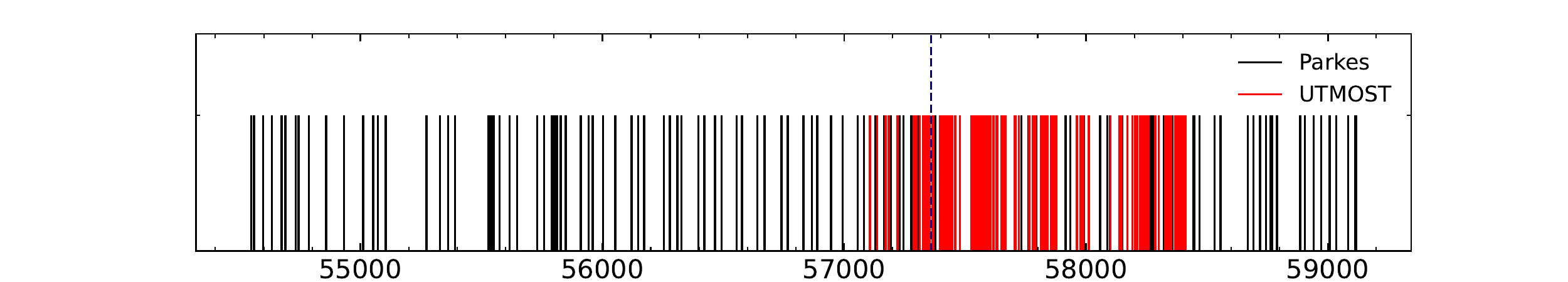}
\caption{Distributions of Parkes (black lines) and UTMOST (red lines) observations for PSR J0738$-$4042. Parkes data have a 332 d observation gap after the measured glitch epoch marked by a blue dashed line.}
\label{fig:observations}
\end{figure*}

PSR J0738$-$4042 has long been become a continuously tracking target at the centre frequencies of 843 MHz and 1369 MHz with the Upgraded Molonglo Observatory Synthesis Telescope (UTMOST) and Parkes 64-m Radio Telescope, respectively. All timing observations obtained for this work are publicly available from the Molonglo Online Repository\footnote{\href{https://github.com/Molonglo/TimingDataRelease1/}{https://github.com/Molonglo/TimingDataRelease1/}} \citep{lbs20} and Parkes Pulsar Data Archive\footnote{\href{https://data.csiro.au/dap/public/atnf/pulsarSearch.zul}{https://data.csiro.au/dap/public/atnf/pulsarSearch.zul}} \citep{hmm11}. Fig. \ref{fig:observations} shows the distribution of the collected 12.5-yr observations ranging from MJDs 54548 (March 2008) to 59116 (September 2020) by Parkes and UTMOST in detail. Here, Parkes data have a 332 d observation gap after the measured glitch epoch so that only UTMOST data were used for determining glitch parameters. The extant UTMOST ToAs together with Parkes data have been used to analyse the long-term rotational properties whereas for pulse profile variations only Parkes data have been processed. The details of the timing observations for this pulsar conducted at UTMOST and Parkes observatory are described in \cite{jbs19} and \cite{ljd21}. Meanwhile, the HartRAO ToAs recorded between MJDs 45704 (January 1984) and 54743 (October 2008) in the work of \cite{bkb14} are obtained to revisit a previous mode-switching behaviour.

Pulsar timing techniques involve the evaluation and interpretation of the differences (also known as ``timing residuals'') between the observed pulse times-of-arrival (ToAs) and predicted arrival times, which help to examine the entire rotational history of a given pulsar. In order to determine the actual ToAs, the software packages \textsc{psrchive} \citep{hvm04} are employed in these procedures. After removing radio-frequency interference (RFI) and being de-dispersed, each observation is scrunched in time, frequency and polarisation to extract mean pulse profiles, which are subsequently aligned and added to form a standard profile template. After that the \texttt{pat} tool of \textsc{psrchive} is invoked to perform  a cross-correlation between the observed mean profiles and the standard template to yield highly accurate ToAs of topocentric pulse. Next, these ToAs need to be transformed to the Solar system barycentre (SSB) based on the Jet Propulsion Laboratory's (JPL) DE430 Solar-system ephemeris \citep{ftb14} and the TCB (Barycentric Coordinate Time) scale.
The high-precision pulsar timing analysis software \textsc{tempo2} \citep{hem06}  is used to obtain the timing residuals with a rotational phase model, which is introduced by a Taylor series expansion \citep{ehm06}:

\begin{equation}\label{euqation_1}
\phi(t)=\phi(t_{0})+\nu(t-t_{0})+\frac{\dot{\nu}}{2}(t-t_{0})^{2}+\frac{\ddot{\nu}}{6}(t-t_{0})^{3}+...,
\end{equation}
where $\phi$, $\nu$, $\dot{\nu}$ and $\ddot{\nu}$ are the phase, spin-frequency and its derivatives as measured at the fiducial epoch $t_{0}$.

Conventionally, a glitch will result from an additional pulse phase modelled by the equation \citep{ehm06}: 
\begin{equation}
\begin{split}
\phi_{\rm g}=&\Delta\phi+\Delta\nu_{\rm p}(t-t_{\rm g})+\frac{1}{2}\Delta\dot\nu_{\rm p}(t-t_{\rm g})^{2}+\\
&[1-e^{-(t-t_{\rm g})/\tau_{\rm d}}]\Delta\nu_{\rm d}\tau_{\rm d},
\end{split}
\label{Eq_decay}
\end{equation}
where $\Delta\phi$ is an offset in the pulse phase at the glitch epoch $t_{\rm g}$. The permanent increments in the pulse frequency $\Delta\nu_{\rm p}$ and first frequency derivative $\Delta\dot{\nu}_{\rm p}$, in addition to a transient frequency increment $\Delta\nu_{\rm d}$ with an exponential decay timescale $\tau_{\rm d}$, are used to model a glitch. Hence, the observed relative glitch sizes in the rotation and the spin-down rates are described as: 
\begin{equation}
\frac{\Delta\nu}{\nu}=\frac{\Delta\nu_{\rm p}+\Delta\nu_{\rm d}}{\nu},
\end{equation}
\begin{equation}
\frac{\Delta\dot{\nu}}{\dot{\nu}}=\frac{\Delta\dot{\nu}_{\rm p}-\Delta\nu_{\rm d}/\tau_{\rm d}}{\dot{\nu}}.
\end{equation}
Furthermore, the parameter $Q$, which is defined as the ratio $\Delta\nu_{\rm d}/\Delta\nu$, quantifies the fraction of glitch recovery.


\begin{figure*}
\centering
\includegraphics[width=0.48\textwidth]{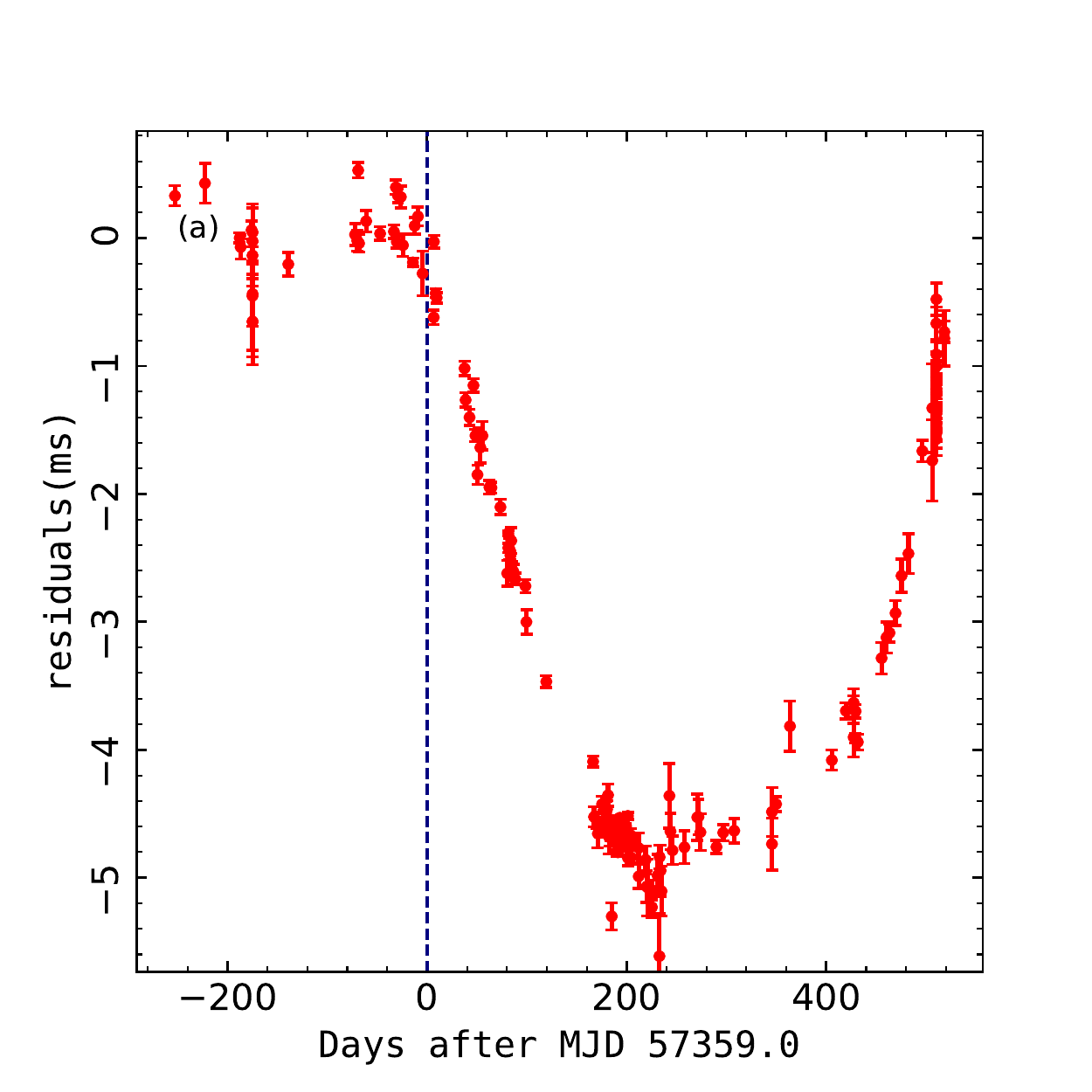}
\includegraphics[width=0.48\textwidth]{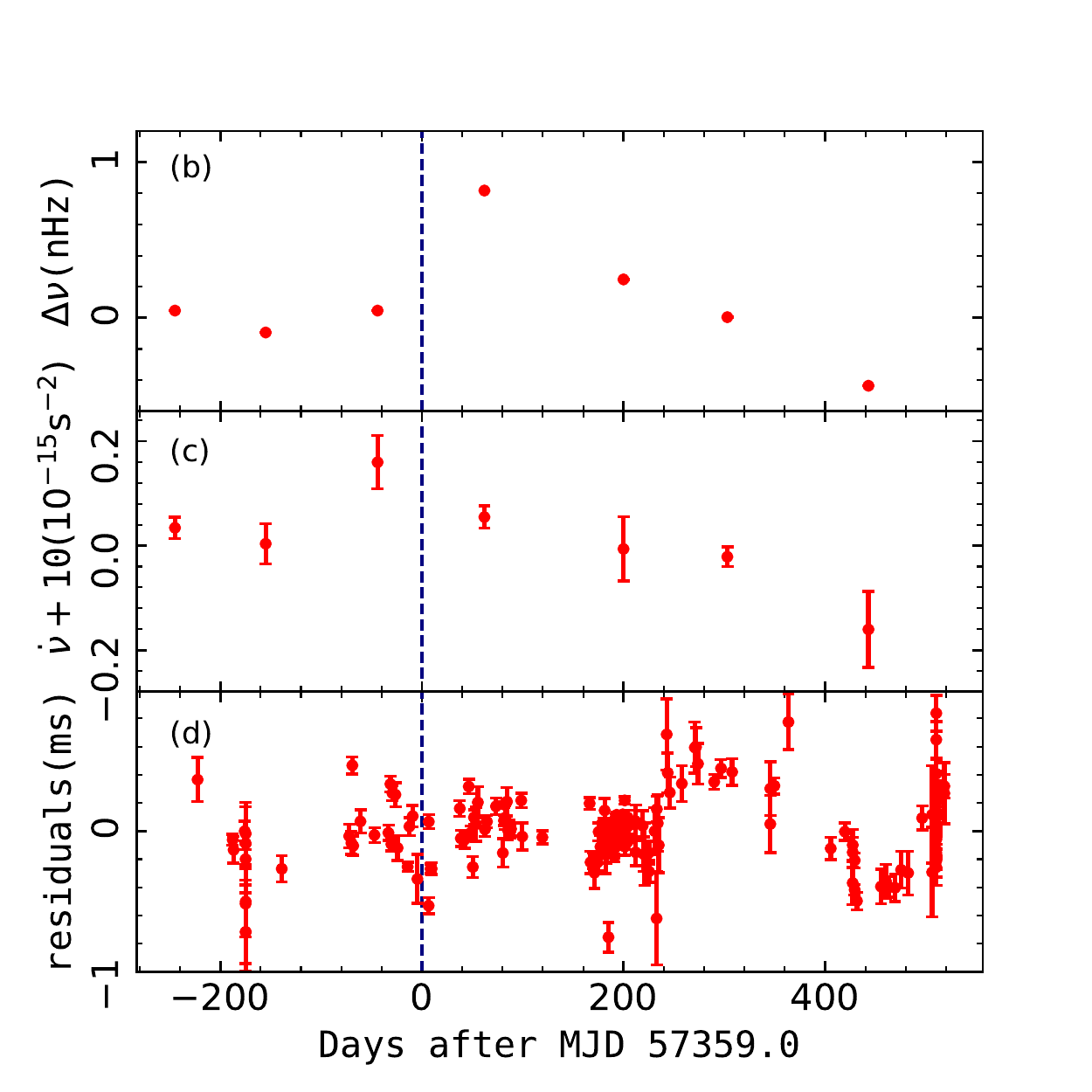}
\caption{First glitch observed in PSR J0738$-$4042: (a) timing residuals with respect to the pre-glitch spin-down model; (b) variations of the frequency resuduals $\nu$ after subtracting the pre-glitch spin-down model; (c) the variations of $\dot{\nu}$; (d) the post-fit residuals after adding the glitch terms listed in Table \ref{tab:glitch} to the timing model. This glitch analysis is conducted with only UTMOST data between MJDs 57106 and 57878.
The vertical lines indicate the glitch epoch at MJD $\sim57359(5)$.}
\label{fig:glitch}
\end{figure*}


\begin{table}
\label{tab:glitch}
\caption{The fitted timing solutions and glitch parameters.}
\begin{center}
\begin{threeparttable}
\begin{tabular}{lccc}

\hline
\hline  
Parameter                                           &       Values               \\
\hline
$\nu$ (Hz)                                          &      2.6672305326(2)       \\
$\dot{\nu}$ ($10^{-15}$\rm\ s$^{-2}$)               &      --9.73(1)             \\
Freq. epoch (MJD)                                   &      57491                 \\
TOA numbers                                         &       177                  \\
Data range (MJD)                                    &    57106--57878            \\
Rms residual ($\mu {\rm s}$)                        &      578$^*$/116$^*$       \\
Glitch epoch (MJD)                                  &       57359(5)              \\
$\Delta\nu$ ($10^{-10}$\rm\ Hz)                     &        9(1)                \\
$\Delta\nu/\nu$ ($10^{-9}$)                         &       0.36(4)              \\
$\Delta\dot{\nu}$ ($10^{-17}$\rm\ s$^{-2}$)         &       --3(1)               \\
$\Delta\dot{\nu}/\dot{\nu}$ ($10^{-3}$)             &        3(1)                \\
\hline

\end{tabular}
\begin{tablenotes}
\item[] *: Rms before and after fitting the glitch.
\end{tablenotes}
\end{threeparttable}
\end{center}
\end{table}

\begin{table*}
\caption{PSR J0738$-$4042's timing solutions before and after the glitch, produced by TEMPO2.}
\begin{center}
\begin{threeparttable}
\begin{tabular}{lccc}

\hline
\hline  
\multirow{2}{*}{Parameter}                     &\multicolumn{2}{c}{MJD range}           &                            \\
                                               &        54548--57355                    &       57363--59116         \\
\hline
Pulsar name                                    &\multicolumn{2}{c}{PSR J0738$-$4042}     &                            \\
Right ascension, RA (J2000) (h:m:s)            &\multicolumn{2}{c}{07:38:32.244(3)}     &                            \\
Declination, DEC (J2000) (d:m:s)               &\multicolumn{2}{c}{--40:42:39.43(4)}    &                            \\
Pulse frequency, $\nu$ (Hz)                    &2.66723182723(3)                        &2.66722989447(2)            \\
First derivative of pulse frequency, $\dot{\nu}$ ($10^{-15}$\rm\ s$^{-2}$)
                                               &--9.7666(4)                             &--9.9875(4)                 \\
Second derivative of pulse frequency, $\ddot{\nu}$ ($10^{-24}$\rm\ s$^{-3}$)
                                               &0.63(1)                                 &--2.89(3)                   \\
Frequency epoch (MJD)                          &55951                                   &58239                       \\
Number of ToAs                                 &117                                     &265                         \\
RMS timing residual ($\mu {\rm s}$)            &3601                                    &2247                        \\
Dispersion measure, DM ($\rm cm^{-3}\ pc$)     &\multicolumn{2}{c}{160.8(6)}            &                             \\
Time--scale                                    &\multicolumn{2}{c}{TCB}                 &                            \\
Solar system ephemeris model                   &\multicolumn{2}{c}{DE430}               &                            \\
\hline

\end{tabular}
\end{threeparttable}
\end{center}
\label{tab:solutions}
\end{table*}


\section{Glitch}\label{glitch}
In order to analyse the long-term spin behaviour of PSR J0738$-$4042, the weighted least-square fit is utilized to find a set of phase-connected timing solutions. The pulsar's timing position ($\alpha$, $\delta$) referred to the result updated by \cite{lbs20} and a constant phase offset between the UTMOST and Parkes ToAs, are included to determine a good timing model. In this process, a signature of the presence of one possible typical small glitch is identified. As is shown in panel (a) of Fig. \ref{fig:glitch}, the timing residuals relative to the spin-down model before MJD $\sim 57359(5)$ are randomly distributed around zero and then show a downward trend with a parabola-like structure. To confirm if a small amplitude jump in the spin frequency has actually occurred, the time evolution of $\nu$ and $\dot{\nu}$, obtained from separate fits to a short set of overlapping observations, are displayed in panels (b) and (c) of Fig. \ref{fig:glitch}. The frequency, $\nu$, varies by $0.9(1)\times10^{-9}\rm \ Hz$, and there is a slight decrease in the frequency derivative , $\dot{\nu}$, of $3(1)\times10^{-17}\rm \ s^{-2}$. Apart from this, the pattern of the post-fit residuals after adding the glitch terms to the timing model in panel (d) of Fig. \ref{fig:glitch} indicates that glitch parameters have been properly modelled. Taking all these together, a small glitch is verified to occur at MJD $\sim 57359(5)$. Glitch parameters are obtained by fitting equation (\ref{Eq_decay}) without the exponential decay as there is no clear evidence for this term. Because the glitch epoch cannot be determined from the \textsc{tempo2} timing fit, it was set at the middle date of the segment between the last pre-glitch observation and the first post-glitch observation, with an uncertainty of half the observation gap. Our fitted values presented in Table \ref{tab:glitch} give $\Delta\nu/\nu = 0.36(4)\times10^{-9}$ and $\Delta\dot{\nu}/\dot{\nu} = 3(1)\times10^{-3}$. 

Timing solutions for the pre- and post-glitch data spans are given in Table \ref{tab:solutions}. These parameter values before and after the glitch are generally consistent with the ones in \cite{bkj16} and \cite{ljd21}, respectively. We notice that this pulsar displays an apparent reversal of the sign of $\ddot{\nu}$ at the time of the glitch, with a significant difference ($\Delta\ddot{\nu} = -3.52(3)\times10^{-24}\rm \ s^{-3} $). The underlying reason is confidently the unusual post-glitch behaviour in this pulsar, since almost all glitches display both a post-glitch value of $\ddot{\nu} \geq 0$ and a difference of $\Delta\ddot{\nu} \geq 0$ \citep{mh11}. In the present case, the whole 12.5-yr data set is binned into intervals of $\sim 200$ days to fit separately with the standard spin-down model to demonstrate in depth the evolution of $\dot{\nu}$ in panel (a) of Fig. \ref{fig:profile}. A permanent change in $\ddot{\nu}$ is confirmed with the very different slopes of $\dot{\nu}$ before and after the glitch. Clearly, this pulsar behaves in a new spin-down state after the glitch. The $|\dot{\nu}|$ steadily increases over $\sim 1380\ \rm d$ to a peak of $10.11(1)\times10^{-15}\ \rm s^{-2}$, and thereafter shows a decay trend. Quoting the observable $\nu$ and its first two derivatives in the two independent spin-down states from Table \ref{tab:solutions}, we obtain the pre- and post-glitch braking indices ($\ddot{\nu}\nu/{\dot{\nu}}^{2}$) $n_{\rm pre} = 17630(580)$ and $n_{\rm post} = -77410(880)$, the errors calculated with the linear error propagation method. \cite{ljd21} obtained a braking index value of $-96227$ for PSR J0738$-$4042 by using a much shorter 3.6-yr UTMOST data. According to \cite{jg99} and \cite{ab06}, a possible reason for such large braking indices is the unresolved effects of the glitches and the superfluid internal torques in the post-glitch relaxation process. \cite{hlk10} and \cite{dym20} suggested that red timing noise should be recognized as the cause of the middle-age pulsars' anomalously large braking indices. \cite{yz15} modeled that the fluctuation of the magnetic inclination angle $\alpha$, which would correspond to a change in the effective emission geometry \citep{kyw18}, contributes a source of red timing noise. Prior to this work \cite{lbs20} explored the timing noise properties of the 300 UTMOST-observed radio pulsars and found that PSR J0738$-$4042 is the only pulsar which has been shown to be in favour of the power-law red noise with $\ddot{\nu}$ (PLRN+F2) model. Alongside this, \cite{lbs20} speculated this case could be due to not considering the state-changing behaviours in their timing noise model. These observations mean that the following analysis of the evolution of the pulse profile is an essential ingredient for characterising the true spin evolution of PSR J0738$-$4042.

\section{Pulse Profile Changes Correlated with spin-down changes}
\begin{figure*}
\centering
\includegraphics[width=0.48\textwidth]{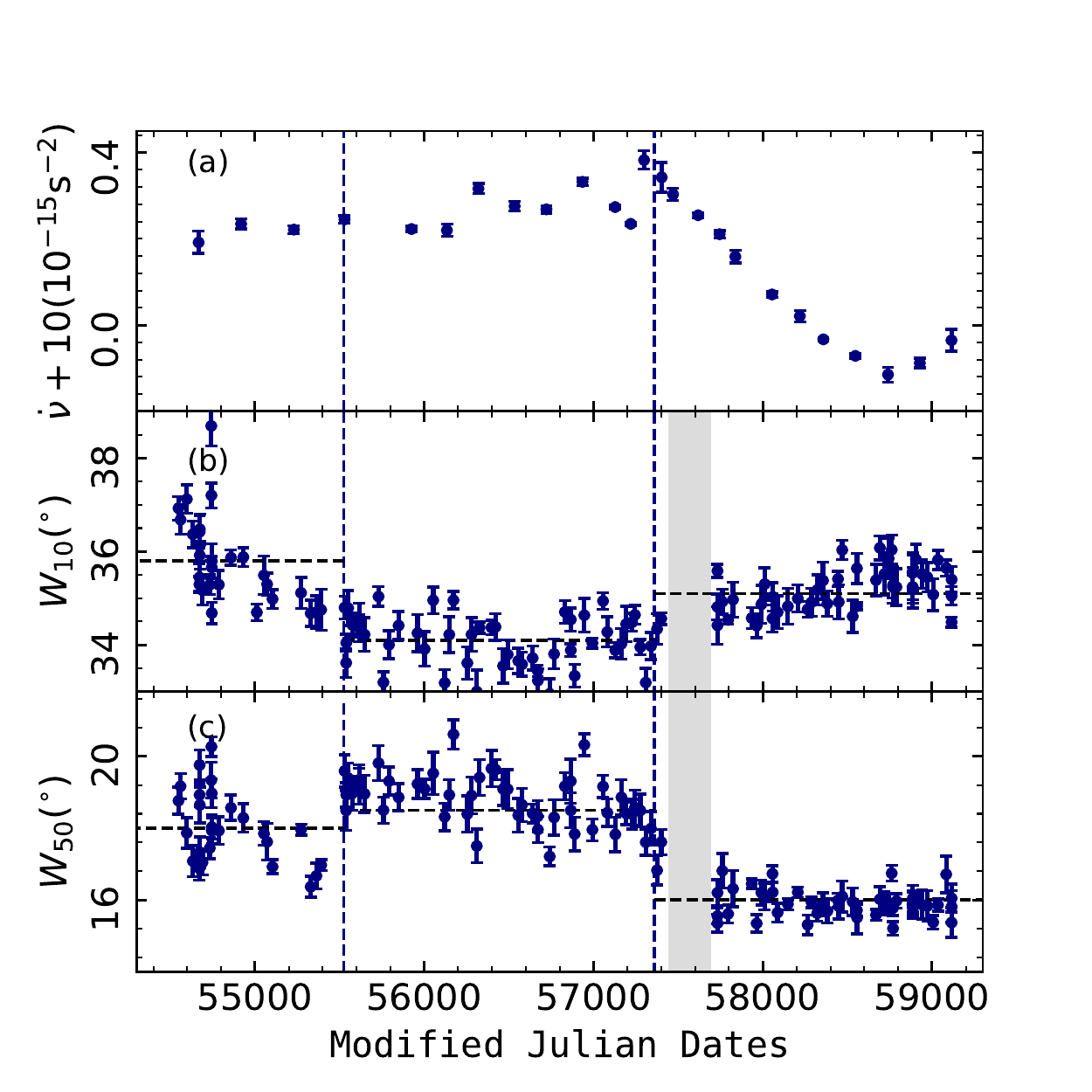}
\includegraphics[width=0.48\textwidth]{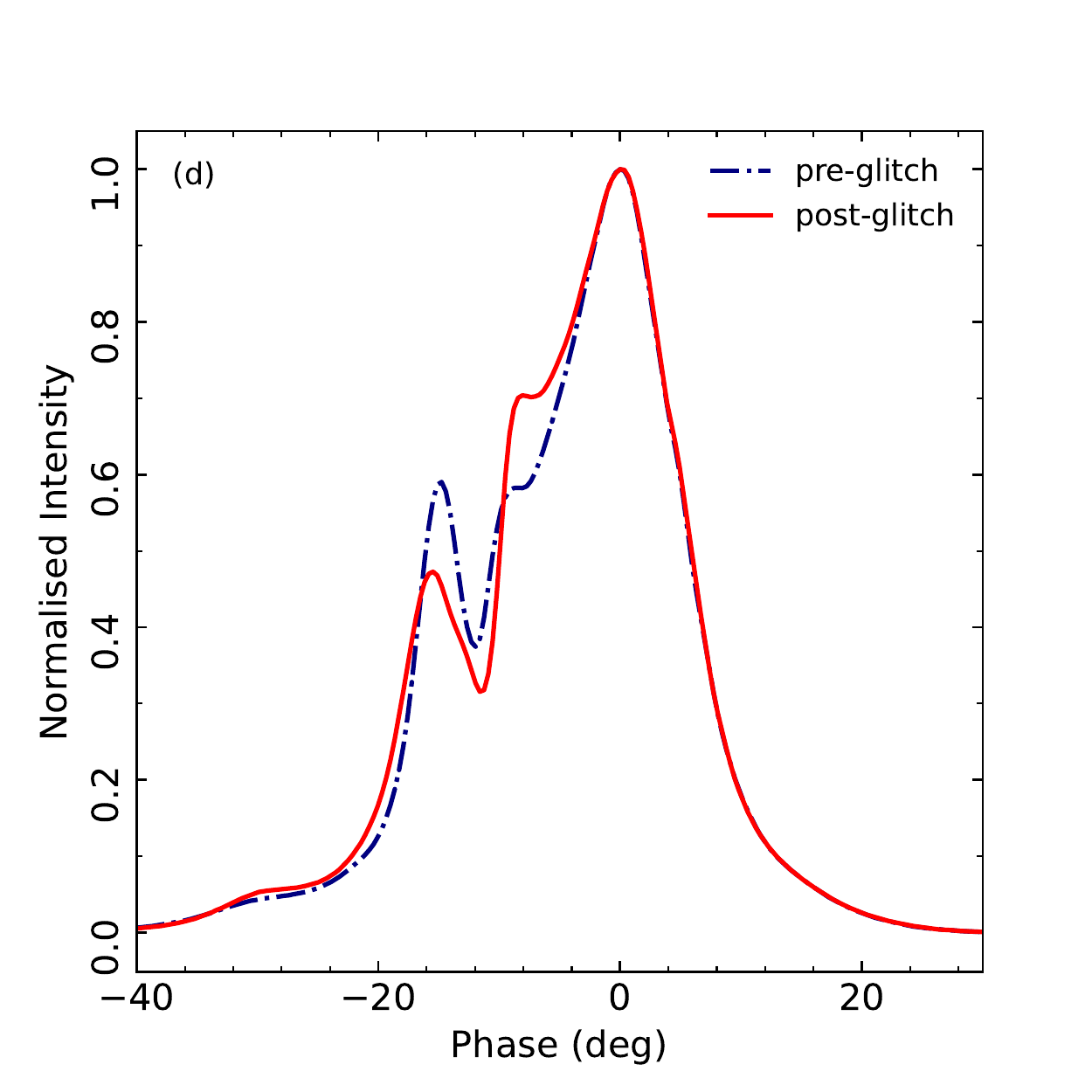}
\caption{Emission-rotation correlation in PSR J0738$-$4042: (a) long-term evolution for the spin-down rate $\dot{\nu}$ obtained from  the combined UTMOST and Parkes data; (b) and (c) the pulse width $W_{\rm 10}$ and $W_{\rm 50}$ at $10\%$ and $50\%$ of the peak for all 1369-MHz integrated pulse profiles observed at Parkes, respectively; (d) the integrated normalized pulse profiles at pre- (blue line) and post-glitch (red line) modes. Grey areas in panels (b) and (c) signify the observation gap between MJDs 57400 and 57732 of Parkes data. The two vertical lines mark the epoch at MJD $\sim55525$ and the first glitch occurred at MJD $\sim57359(5)$. The horizontal lines
are stand for the average values of $W_{\rm 10}$ and $W_{\rm 50}$ (listed in Table \ref{tab:w50}) at three different modes.}
\label{fig:profile}
\end{figure*}

{\label{correlation}}
PSR J0738$-$4042 underwent a switching between two- and three-components of the pulse profile since its first spin-down state transition event occurred in September 2005 \citep{bkb14}. The triple-peaked profile of PSR J0738$-$4042 contains a dominant trailing component, preceded by two obvious shoulder peaks \citep{bkj16}. In Panels (b) and (c) of Fig. \ref{fig:profile} we plot the pulse width at the 10 percent level of the pulse peak $W_{\rm 10}$ and full width at half maximum of the peak $W_{\rm 50}$ to present the evolution of the average profiles taken with the 20-cm receiver of the Parkes telescope at various times of the last 12.5-yr. Visually, the widths of both types has changed right after MJD $\sim 55525$, but the spin-down rates in Panel (a) of Fig. \ref{fig:profile} are not observed to exhibit any unusual behaviour around this time. \cite{bkj16} also reported this drastic variability in the pulse shapes. Since then, the mean pulse profiles maintain the new relatively stable state until the occurrence of this glitch, after which the profile changes get involved. It is impossible to determine whether the pulse width change has taken place suddenly or gradually after the time of the glitch due to missing coverage of observations, but it is still clear that $W_{\rm 10}$ increases while $W_{\rm 50}$ decreases followed by a steady value. Table \ref{tab:w50} provides the average pulse width measurements for each group of the observed profiles. The average $W_{\rm 50}$ values are $18.5(4)^{\circ}$ before and $16.0(3)^{\circ}$ after the glitch. To illustrate the specific changes in the pulse profile, normalized integrated pulse profiles before and after glitch, which are derived from summing all aligned pulse profiles in MJDs 55529--57355 and MJDs 57363--59116 respectively, are plotted in panel (d) of Fig. \ref{fig:profile}. The leading peak of the post-glitch normalized profile becomes weaker, contrasting with its pre-glitch relative size, and is accompanied by an enhancement in the central component. It is worthy to note that this pulsar has a slightly wider mean pulse profile after the glitch. These concurrent changes in the average pulse profile and spin-down rate state suggest that the origin of the mode-switching phenomenon in PSR J0738$-$4042 is related to the glitch event.


\begin{table*}
\label{tab:w50}
\caption{Pulse widths $W_{10}$ and $W_{50}$ in three different spin-down states.}
\begin{center}
\begin{threeparttable}
\begin{tabular}{lccc}

\hline
\hline  
Parameter                           & Pre-glitch               & Pre-glitch              & Post-glitch    \\
\hline
Data span (MJD)                     &     54548--55391         &      55529--57355        & 57363--59116    \\
Mean $W_{10}(^{\circ})$             &       35.8(2)            &        34.1(2)           & 35.1(2)          \\
Mean $W_{50}(^{\circ})$             &       18.0(3)            &        18.5(4)           &  16.0(3)        \\
Number of pulse profiles            &         23               &          54              &  51               \\

\hline

\end{tabular}
\end{threeparttable}
\end{center}
\end{table*}


\section{The Combined Model of Crustquake and Vortex Creep Response for Pulsar Glitches and Pulse Shape Change}\label{model}

Vortex lines in the inner crust and core superfluids of a neutron star strongly interact with the lattice nuclei and magnetic flux tubes, respectively, and maintain the rotational equilibrium between the observed crust and interior superfluid components by thermal activation, a process called creep \citep{alpar84,erbil14}. The steady-state creep process can be affected either by rotational or thermal perturbations \citep{cheng88,link96}. Glitches as sudden and collective vortex unpinning events temporarily stop the creep rate within the glitch affected superfluid regions and the post-glitch recovery reflects the tendency to return back to the pre-glitch conditions.

Crustquakes are invoked to explain smaller Crab-like glitches \citep{baym71,reisenegger21} and may play a dominant role in unpinning trigger for larger glitches \citep{ruderman76,alpar96} since a single vortex line freed by a quake is able to release a large number of vortices in high density traps \citep{cheng88,warszawski12}. There are three main contributions to strain in the crust that generate a fracturing quake: a combination of gravitational force and centrifugal force with secular spin-down \citep{giliberti20,reisenegger21} or crustal magnetic field \citep{franco00,lander15} or pinning stress due to vortex-nuclei interaction \citep{ruderman76,anderson82} and vortex-flux tube entanglement at the base of the crust \citep{ruderman98,sedrakian99}. The spin-down induced stresses act to reduce the centrifugal bulge of the neutron star crust formed when the star was rotating much faster. Thus, after experiencing successive glitches the neutron star assumes a spherical shape which limits the extent of the crustquakes via spin-down. Given the surface magnetic field strength estimate from the dipole formula is $B_{\rm d}=7.27\times10^{11}$ G for PSR J0738$-$4042, the crustal magnetic field originated stresses are not expected to play an important role in quakes. Since both the spin-down and crustal magnetic field induced stresses are in the direction of the magnetic pole, their contributions will be to decrease inclination angle after a quake \citep{lander15}.     

The stresses associated with vortex line-flux tube pinning at the base of the crust, on the other hand, is in the direction of the neutron star equator \citep{ruderman98,erbil16}. Whenever this continuous growth rate of the stress approaches the yielding point a crust breaking quake takes place and leads broken platelet to displace. The event may induce inward superfluid vortex line motion towards the rotational axis and in turn initiate collective vortex unpinning, thus playing the role of glitch trigger. Since the footpoints of global magnetospheric field lines are anchored in the crust, any irreversible motion of the broken platelet will be accompanied by tiny growth in the inclination angle between the rotation and magnetic dipole moment axes. Another influence of the platelet movement will be the small scale increase of the curvature of the local magnetic field lines in the vicinity of the polar cap, a process that may cause emission changes. The changes in the radio emission arise from the modifications of the surface magnetic field in the inner magnetospheric region above the polar cap \citep{geppert21}. Here is the place where radio emission originates through electron/positron pairs formation and acceleration in a sufficiently strong and curved field ensues. Since the outer gap regions wherein X-ray and gamma radiation presumably occur are located well above the polar cap, it is more difficult to observe pulse profile variations in these short wavelengths concurrent with the glitch event.   

The pulse width at half maximum $W_{50}$ is related to the pulse period $P$ and the inclination angle $\alpha$ via \citep{rankin90,gil11}
\begin{equation}
    W_{50}=2^{\circ}.45 \frac{P^{-1/2}}{\sin\alpha}.
\label{rankinalpha}    
\end{equation}

A similar fit formula applied to a larger sample of radio pulsars exists for $W_{10}$ values \citep{posselt21}
\begin{equation}
W_{10}=(11.9\pm0.4)^{\circ}(P/\mbox{s)}^{-0.63\pm0.06}.
\label{w10}
\end{equation}
The mean $W_{50}$ values in Table \ref{tab:w50} for the pulse width measurements before and after the 2015 glitch event yield, by using equation (\ref{rankinalpha}), the inclination angles $12^{\circ}.5$ and $14^{\circ}.6$ respectively, before and after the glitch of PSR J0738$-$4042. Therefore, the increment of $\Delta\alpha=2^{\circ}.1$ in the inclination angle seems to be produced by the glitch event itself. However, $W_{10}$ value predicted by equation (\ref{w10}) $(20^{\circ}-24^{\circ})$ differs significantly from the observed $W_{10}\approx35^{\circ}$, indicating that newly emerged component affected the pulse profile largely.

In the plasma filled magnetosphere inclination angle evolution under the the central dipole radiation approximation leads to the following equation which is constant throughout the lifetime of a pulsar \citep{philippov14,eksi16} 
\begin{equation}
    C\equiv\frac{\cos^2{\alpha_{0}}}{P_{0}\sin{\alpha_{0}}}=\frac{\cos^2{\alpha}}{P\sin{\alpha}},
\label{mhdpulsar}    
\end{equation}
where the subscript `0' refers to the measurement of the corresponding quantity at a particular time. The plasma magneto-hydrodynamics pulsar alignment time-scale for PSR J0738$-$4042 is $\tau^{MHD}_{\rm align}=\tau_{\rm c}\sin^2{\alpha_{0}}/\cos^4{\alpha_{0}}=0.3$ Myr which means that pulsar will evolve toward alignment on a long time-scale provided that its evolution is not interrupted by other timing events.  

Another effect of the crust breaking event on the pulsar dynamics would be its imprints on the rotational evolution. Crust breaking may be an agent for variation of the steady-state vortex current in the inner crust of neutron stars, in particular play an important role in glitches by transporting some of the vortices inward and unpinning the others. 
This model was successfully applied to the timing behaviours of PSR J1119--6127 \citep{akbal15} and the Crab pulsar \citep{erbil19}.    
Our model for the post-glitch spin-down rate evolution is \citep{erbil19}
\begin{align}
\Delta\dot\nu (t)=
&-\frac{I_{\rm A'}}{I}\left|\dot\nu_{\infty}\right|\left[1-\frac{1}{1+\left(e^{-t'_{0}/\tau'_{\rm nl}}-1\right)e^{-t/\tau'_{\rm nl}}}\right]
\nonumber\\
&-\frac{I_{\rm A}}{I}\left|\dot\nu_{\infty}\right|\left[1-\frac{1}{1+\left(e^{t_{0}/\tau_{\rm nl}}-1\right)e^{-t/\tau_{\rm nl}}}\right],
\label{creep}
\end{align}
where $\dot\nu_{\infty}$ is the spin-down rate just before the glitch, $I_{\rm A}/I$, $t_{0}$, $\tau_{\rm nl}$ are nonlinear superfluid region's fractional moment of inertia, offset time, and recoupling time-scale for outward moving vortex lines at the time of glitch and primed quantities are their inward moving vortices counterparts. The nonlinear creep regions are responsible for glitches since through these regions conditions are very close to lead to vortex unpinning avalanche. The post-glitch response of these nonlinear regions are determined by two timescales: recoupling time-scale $\tau_{\rm nl}$ is the time for coupling of the glitch affected nonlinear superfluid region to the other superfluid components of the star. The offset time $t_{0}$ determines the time-scale on which glitch affected decrease in the angular velocity lag between the superfluid and crustal rotational rates turns back towards the original pre-glitch value again by the neutron star spin-down. Note that if equal number of vortex lines move inward and outward after the collective unpinning event and traverse the same radial extent, then glitch magnitude in the spin frequency will be tiny or even zero. However, due to the time variation of the interior superfluid torque acting on the neutron star, the spin-down rate will first experience a gradual slow increase, i.e. become more negative, as a result of inward motion of vortices, and after a certain time it will reverse and start to decay toward the original pre-glitch value as outward moving vortex lines repin and restart creep again.       

\begin{figure}
\centering
\vspace{0.1cm}
\includegraphics[width=1.0\linewidth]{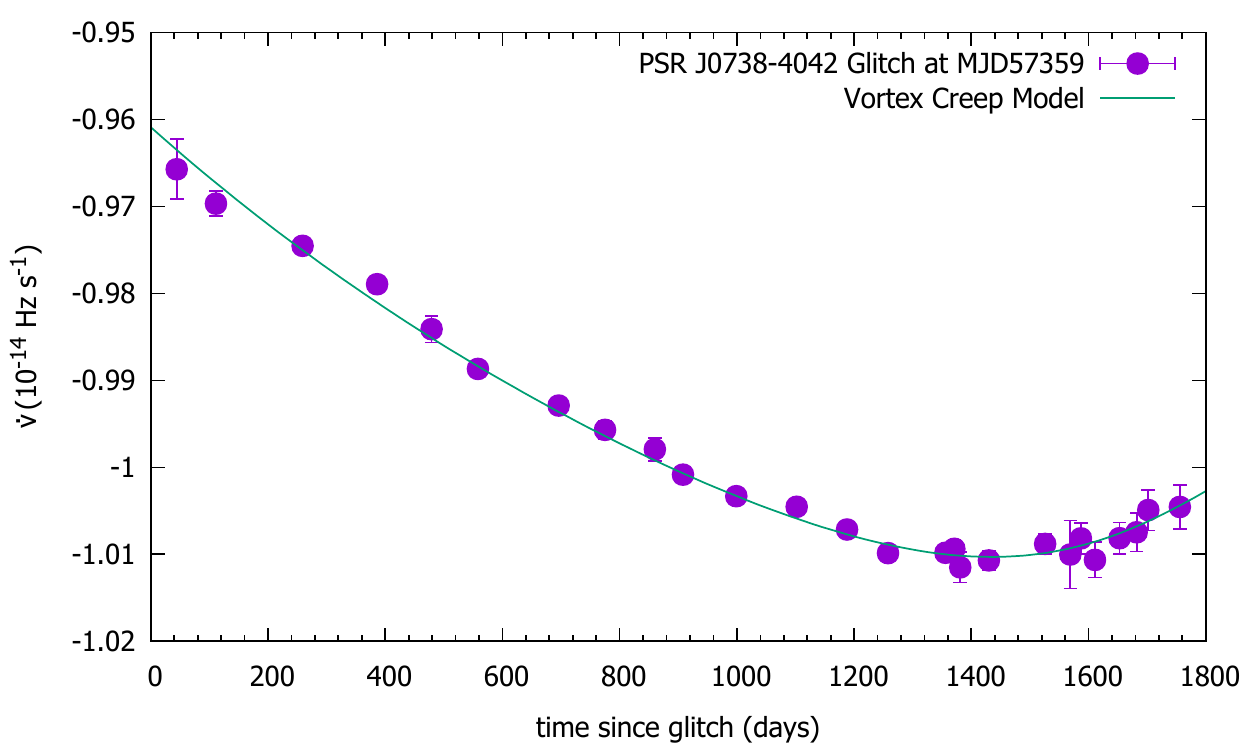}
\caption{Comparison between the glitch occurred at MJD 57359 (purple) and the vortex creep model prediction (green).}
\label{modelfit}
\end{figure}

We apply equation (\ref{creep}) to the post-glitch recovery of 2015 event. Comparison between the vortex creep model and the 2015 glitch data is displayed in Fig. \ref{modelfit}. Best fit model parameters are shown in Table \ref{model}.

\begin{table}
\centering
\caption{Inferred vortex creep model parameters for the fit to PSR J0738$-$4042 post-glitch timing data. Here have used the standard notation that $(A)_{\rm a}=A\times10^{\rm a}$.}
\begin{tabular}{ll}
\hline\hline
Parameter & Value \\
\hline
$\left(\frac{I_{\rm A'}}{I}\right) _{-2}$ & 1.9(2) \\
$\left(\frac{I_{\rm A}}{I}\right) _{-2}$ & 4.9(6) \\
$\left(\frac{I_{\rm B}}{I}\right) _{-2}$ & 1.6(7) \\
$ t'_{0} (\rm days) $ & 41238(4578) \\
$ t_{0}(\rm days) $ & 1868(310) \\
$ \tau'_{\rm nl} (\rm days) $ & 19600(1902) \\
$ \tau_{\rm nl} (\rm days) $ & 211(19) \\
$ \delta\Omega'_{\rm s} (\mbox{\rads})_{-6}$ & 37.5(5.0) \\
$ \delta\Omega_{\rm s} (\mbox{\rads})_{-6}$ & 1.7(3) \\
\hline
\label{model}
\end{tabular}
\end{table} 

Given that non-linear superfluid relaxation time-scale is $\tau'_{\rm nl}\approx50$ yr, the spin-down rate fluctuations before the glitch can be understood as time variable superfluid coupling to the external pulsar braking torque \citep{erbil17}. Note that damped sinusoidal-like oscillations had also been observed before and after the 1988 Christmas glitch of the Vela pulsar \citep{mcculloch90} and second largest ever glitch in PSR B2334+61 \citep{yuan10}. The fractional moment of inertia of the non-linear creep regions participated to PSR J0738$-$4042 glitch is similar to those of Vela-like pulsars \citep{egy22}, in line with the expectation that older pulsars have well-established vortex trap network \citep{ab06}. The total number of vortices involved in the 2015 glitch can be estimated as
\begin{equation}
N_{\rm v}=\frac{2\pi R^{2}\left(\delta\Omega_{\rm s}+\delta\Omega'_{\rm s}\right)}{\kappa}= 5.43\times10^{5}\frac{(t_{0}+t'_{0})|\dot\nu_{\infty}|R^{2}}{\kappa}, 
\end{equation}
with $\kappa=2\times10^{-3}$\qvorticity~ being the quantized vorticity attached to each vortex line and $R$ being the neutron star radius and the change in the superfluid angular velocity $\delta\Omega_{\rm s}$ due to vortex discharge is related to offset time $t_{0}$ with $\delta\Omega_{\rm s}=2\pi|\dot\nu|t_{0}$. Numerical values in Table \ref{model} gives $N_{\rm v}=3.57\times10^{10}$ which is three orders of magnitude smaller than the values inferred for larger glitches in the Crab \citep{erbil19}, Vela \citep{akbal17,erbil20}, and PSR J1119--6127 \citep{akbal15}. The total fractional crustal superfluid moment of inertia included in the 2015 glitch of PSR J0738$-$4042 from Table \ref{model} is $I_{\rm cs}/I=(I_{\rm A}+I_{\rm A'}+I_{\rm B})/I=(8.4\pm0.9)\times10^{-2}$, which puts a lower bound on the crustal moment of inertia and in turn implies a stiff equation of state for neutron star matter. Here $I_{\rm B}$ is the capacitive vortex traps in which vortices do not creep at all and obtained from the angular momentum balance during the glitch [see, for instance, Equation (12) in \citep{akbal15}]. This $I_{\rm cs}$ is consistent with the fact that neutron stars should have a large crust to deposit substantial amount of angular momentum which will be tapped at giant glitches \citep{delsate16,li21}. 

The total crustal superfluid moment of inertia involved in the 2015 glitch $I_{\rm cs}$ can be used to put a constraint on the surface temperature of PSR J0738-4042. For older pulsars in which the original heat content had been radiated away via neutrino emission and photons from the surface, the residual heat arising from the superfluid friction implies a surface temperature \citep{alpar84}
\begin{equation}
    T_{\rm s}=\left(\frac{I_{\rm cs}\overline{\omega}_{\rm cr}|\dot\Omega|}{4\pi\sigma R^{2}}\right)^{1/4},
\label{surfacetemp}    
\end{equation}
where $\overline{\omega}_{\rm cr}$ is the value of the critical angular velocity between the inner crust superfluid and the normal matter before vortex line unpinning that averaged over the pinning layers, $|\dot\Omega|=2\pi|\dot\nu|$ is the magnitude of the first time derivative of the pulsar angular velocity,  $\sigma$ is the Stefan-Boltzmann constant, $R\cong10^{6}$ cm is the neutron star radius. Since there are other heating mechanisms for older pulsars, like rotochemical heating \citep{reisenegger15} and dissipation due to vortex line-flux tube interaction \citep{sedrakian93}, equation (\ref{surfacetemp}) provides a lower limit to the surface temperature of PSR J0738$-$4042. By taking $\overline{\omega}_{\rm cr}=3.9\times10^{-2}$\rads ~from Table \ref{pinpar} we obtain the estimate $T_{\rm s}=1.38\times10^{5}$ K. Future spectral observations of PSR J0738$-$4042 will reveal the surface blackbody temperature of this middle-aged pulsar. 

\begin{table}
\caption{Vortex line-nucleus pinning related microphysical parameters and nonlinear recoupling time-scale estimates for five crustal layers. Entries in the first three columns are taken from \citet{seveso16} ($\beta=3$ model). The last column is calculated from equation (\ref{sfrecoupling}).}
\label{pinpar}
\begin{center}{
\begin{tabular}{cccc}
\hline\hline\\
\multicolumn{1}{c}{$\rho$} & \multicolumn{1}{c}{$E_{\rm p}$} & \multicolumn{1}{c}{$\omega_{\rm cr}$} & \multicolumn{1}{c}{$\tau_{\rm nl}$} \\
\multicolumn{1}{c}{($10^{13}\mbox{\gcc}$)} & \multicolumn{1}{c}{(MeV)} & \multicolumn{1}{c}{($10^{-2}\mbox{\rads}$)} & \multicolumn{1}{c}{($\mbox{days}$)} \\ 
\hline\\
0.15 & 0.17 & 9.45 & 24125 \\\\
0.96 & 0.24 & 1.44 & 2612 \\\\
3.4 & 2.08 & 11.4 & 2373 \\\\
7.8 & 0.69 & 10.8 & 678 \\\\
13 & 0.02 & 0.02 & 463 \\\\
\hline\\
\label{vortextrap}
\end{tabular}}
\end{center}
\end{table}

From the non-linear superfluid recoupling time-scales $\tau'_{\rm nl}$ and $\tau_{\rm nl}$,
\begin{equation}
    \tau'_{\rm nl}, \tau_{\rm nl}=\frac{kT}{E_{\rm p}}\frac{\omega_{\rm cr}}{|\dot\Omega|},
\label{sfrecoupling}    
\end{equation}
with $E_{\rm p}$ being the pinning energy per vortex line-nucleus intersection and $k$ being the Boltzmann constant, one can estimate the location of the glitch trigger for collective unpinning event. Since the ratio $\omega_{\rm cr}/E_{\rm p}$ depends uniquely on the density of the neutron star crust, for a given microscopic neutron star model $\tau_{\rm nl}$ found from fits helps to constrain the density (location) which is responsible for the glitch trigger. We employ SLy4 equation of state for $1.4M_{\odot}$ neutron star \citep{douchin01} and use \citet{gudmundsson83} formula $T_{8}=1.288(T^{4}_{\rm s6}/g_{\rm s14})^{0.455}$ to convert neutron star surface temperature $T_{\rm s}$ from the estimate given in equation (\ref{surfacetemp}) to internal temperature $T$. Here we define $Q{\rm x} = Q/10^{{\rm x}}$ for the corresponding quantity $Q$ in cgs units. For the chosen neutron star model the gravitational redshift corrected surface gravity is $g_{\rm s14}=1.78$. The non-linear superfluid recoupling time-scales calculated by equation (\ref{sfrecoupling}) for the five distinct pinning layers in the inner crust are displayed in the last column of Table \ref{pinpar}. The non-linear superfluid recoupling time-scale decreases monotonically with increasing density throughout the inner crust. An inspection of Tables \ref{model} and \ref{pinpar} reveals that the vortex inward motion has started from a superfluid layer just beneath of the neutron star outer crust close to the neutron drip point (in connection with the fit result $\tau'_{\rm nl}=19600\pm1902$ days) and the vortex unpinning avalanche extended to the densest pinning layer in the vicinity of the crust-core interface (in connection with the fit result $\tau_{\rm nl}=211\pm19$ days). These findings support our scenario that glitch in PSR J0738$-$4042 has occurred as a result of crustquake induced vortex unpinning avalanche and involved the whole crustal superfluid. The smallness of the glitch magnitude is due to the low number of vortex lines participated to the event.       

The $I_{\rm A'}/I$ value obtained from the post-glitch timing fits can be used to determine the size of the broken platelet. If one assumes a neutron star model in which a thin crust is floating above an incompressible core, $I_{\rm A'}/I$ is related to the broken platelet size as \citep{akbal15,erbil19}  
\begin{equation}
\frac{I_{\rm A'}}{I}\cong \frac{15}{2}\frac{D}{R}\sin{\alpha} \cos^{2}{\alpha}.
\label{plateletsize}
\end{equation}
By using $I_{\rm A'}/I\cong1.9\times10^{-2}$ from Table \ref{model} and $\alpha=12^{\circ}.5$ corresponding to $W_{50}$ value in Table \ref{tab:w50} we obtain $D\approx6\times10^{-4}R=6$ m for the platelet size of PSR J0738$-$4042. This platelet size is similar to the $D=(5.7-18.1)$ m value deduced for the largest glitch in the young Crab pulsar \citep{erbil19}, indicating a universal global pulsar quantity which depends on the neutron star matter properties and the critical strain angle.   

\begin{table}
\centering
\caption{Inferred vortex creep model parameters for the fit to 2007 timing event observed in PSR J0738$-$4042.}
\begin{tabular}{ll}
\hline\hline
Parameter & Value \\
\hline
$\left(\frac{I_{A'}}{I}\right) _{-2}$ & 0.65(16) \\
$\left(\frac{I_{A}}{I}\right) _{-2}$ & 13.7(1.3) \\
$ t'_{0} (\rm days) $ & 2290(1150) \\
$ t_{0}(\rm days) $ & 400(12) \\
$ \tau'_{\rm nl} (\rm days) $ & 1003(662) \\
$ \tau_{\rm nl} (\rm days) $ & 78(10) \\
$ \delta\Omega'_{\rm s} (\mbox{\rads})_{-6}$ & 1.95(98) \\
$ \delta\Omega_{\rm s} (\mbox{\rads})_{-6}$ & 0.34(1) \\
$ t_{\rm th}(\rm days) $ & 634(41) \\
$ t_{\rm obs}(\rm days) $ & 667 \\
\hline
\label{2007timing}
\end{tabular}
\end{table}


\begin{figure*}
\centering
\includegraphics[width=1.0\textwidth]{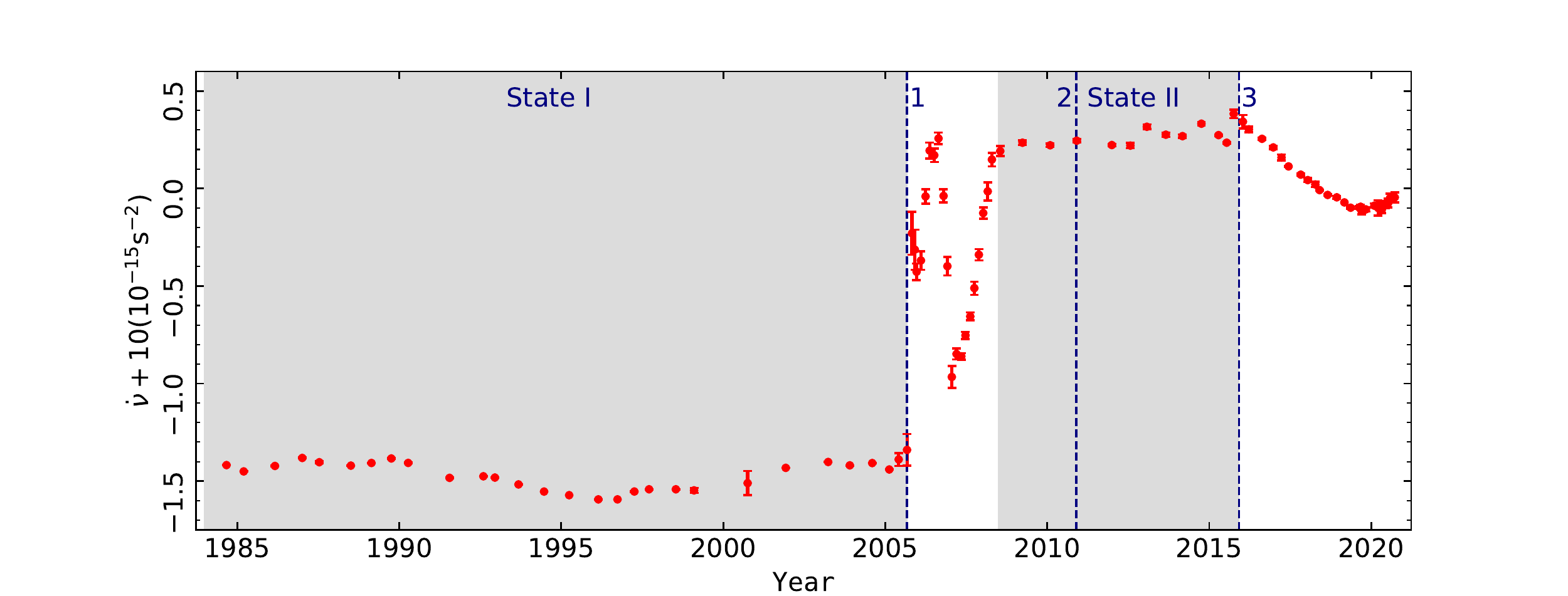}
\caption{The 37-year evolution of the spin-down rate for PSR J0738$-$4042. Two grey areas indicate the high spin-down rate State I ($|\dot{\nu}|_{\rm high} \sim -11.45\times 10^{-15}\ \rm s^{-2}$) and the low spin-down rate State II ($|\dot{\nu}|_{\rm low} \sim -9.76\times 10^{-15}\ \rm s^{-2}$). The vertical dashed lines 1 and 2 indicate the epochs of the pulse profile transitions in September 2005 (MJD $\sim 53614$) and November 2010 (MJD $\sim 55525$), respectively, and line 3 is the 
time that new profile shape changes are associated with a glitch occurred in December 2015 (MJD $\sim 57359$).}
\label{fig:nudot}
\end{figure*}

\begin{figure}
\centering
\vspace{0.1cm}
\includegraphics[width=1.0\linewidth]{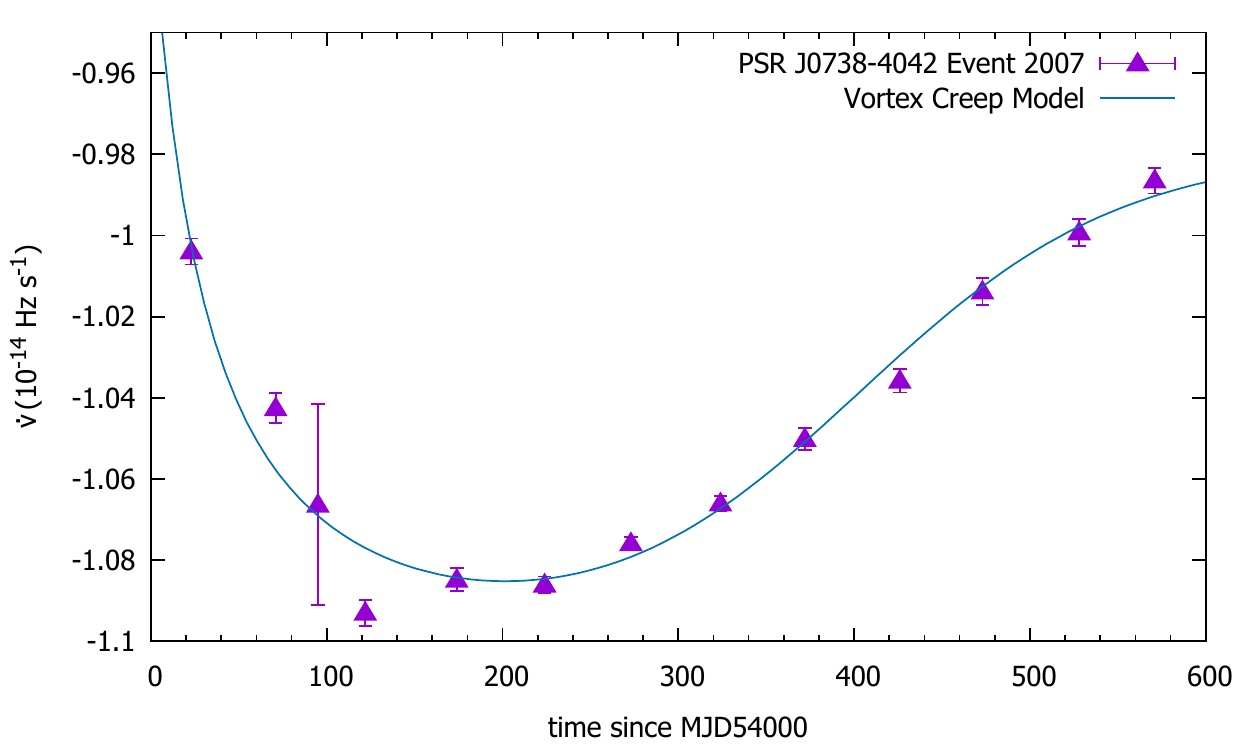}
\caption{Comparison between the timing event occurred at MJD 54000 (purple triangles) and the vortex creep model prediction (blue).}
\label{2007fit}
\end{figure}

We have also applied our model equation (\ref{creep}) to the 2007 timing event observed by \citet{bkb14}. This timing event was preceded by the state transition in the spin-down rate as shown in Fig. \ref{fig:nudot}. The behaviour is reminiscent of the vortex line reconfiguration inside the neutron star just like the case of the 2015 glitch. The fit result is shown in Fig. \ref{2007fit} and the inferred model parameters are given in Table \ref{2007timing}. Theoretical expectation of the duration of the event due to the time variable perturbed interior superfluid coupling with the pulsar braking is $t_{\rm th}=t_{0}+3\tau_{\rm nl}=634(41)$ days and matches quite well with the observed time-scale $t_{\rm obs}=667$ days. The implied radial extent traversed by the unpinned vortices as well as their numbers is slightly larger in this case, presumably due to the involvement of some lines in the core superfluid by thermo-rotational instability. See section \ref{discussion} for details.  

\section{Discussion}{\label{discussion}}
In December 2015 (MJD $\sim 57359$), PSR J0738$-$4042 has experienced a spin frequency jump of magnitude $9(1)\times10^{-10}\rm \ Hz$, that is reported as its first glitch. The discovery of this small glitch, with measured values $\Delta\nu/\nu = 0.36(4)\times 10^{-9}$ and $\Delta\dot{\nu}/\dot{\nu} = 3(1)\times 10^{-3}$, is a serendipitous result of searching for new  variations in the emission profile and $\dot{\nu}$ of this pulsar with the released UTMOST and Parkes timing datasets. This glitch is in the expected size range of the UTMOST undetected events, which was set to $4.1\times 10^{-11} - 2.7\times 10^{-7}$ based on injection studies performed by \cite{dms22}. According to the multiple statistical studies of the glitch behaviours (see, e.g., \cite{dym20}; \cite{bsa22}), a striking aspect is that the observed event sizes of pulsars with large $\tau_{c}$ show a clear trend towards being smaller. For old pulsars with $\tau_{c}>1\ \rm Myr$, $\Delta\nu/\nu$ of almost all glitches are smaller than $10^{-7}$ \citep{lwy21} and their occurrence rate is much lower than that of young pulsars \citep{dym20}. Indeed, PSR J0738$-$4042 is a relatively old pulsar with $\tau_{c}>4\ \rm Myr$ but  underwent only a small glitch in its $\sim 48$-yr of rotational history \citep{Manchester1983,Downs1983,Cordes1985, Downs1986, bkb14}. To quantify a pulsar's glitching rate, \cite{ml90} introduced the glitch activity parameter $A_{\rm g}$, which is defined as $\frac{1}{T}\sum\frac{\Delta\nu}{\nu}$, where $T$ is the total time searched for spin-up events. Compared with glitching pulsars of similar age \citep{dym20}, PSR J0738$-$4042 shows a clearly low value of $A_{\rm g} \sim 7.5(8)\times10^{-12}\ \rm yr^{-1}$.

The spin-down rate of PSR J0738$-$4042 exhibited only a transient small increase at the time of the glitch. The possibility of the existence of an exponential recovery with $\tau_{d} \geq 9 $ days can be safely ruled out, because the data interval around the glitch is less than this timescale. Such behaviours are similar to the other typical small glitches observed to date. Another crucial point for PSR J0738$-$4042 is that, unlike almost all other normal glitches, the post-glitch relaxation has an abnormal $|\dot{\nu}|$ evolution. It is also possible to have large spin-down rate variations for pulsars by magnetospheric charge deviation from the Goldreich-Julian density \citep{shaw22}, asteroid encounter \citep{bkb14,huang16} and internal superfluid modes due to vortex bending \citep{erbil2022} apart from glitches. More specifically, the post-glitch spin-down rate $|\dot{\nu}|$ continues to increase with a parabolic-like tendency. As shown in panel (a) of Fig. \ref{fig:profile}, the post-glitch $|\dot{\nu}|$ is about $5.1(2)\%$ larger than the pre-glitch value, until the turnover point is reached at around MJD 58740 (September 14, 2019). Similar post-glitch features have been seen in some unusual glitches, and can not be understood directly by the standard glitch models. \cite{mh11} detected the largest glitch ever recorded ($\Delta\nu/\nu \sim 33250\times 10^{-9}$) in PSR J1718--3718, with the post-glitch upward trend in $|\dot{\nu}|$ continued over two years. A delayed spin-up of spin frequency in the Crab pulsar during its largest glitch is related with a rapid increase in $|\dot{\nu}|$ \citep{sls18,ge20}. The transition between the dual spin-down states together with the steady increase in $|\dot{\nu}|$ of PSR J1001--5507 is accompanied by pulse shape variations \citep{cb12}, establishing a firm correlation.

Herein, 37-yr evolution of the spin-down rate $|\dot{\nu}|$ for PSR J0738$-$4042 between 1984 and 2021 is revisited and presented in Fig. \ref{fig:nudot}. This pulsar displays mainly two distinct stable spin-down rate (State I: $|\dot{\nu}|_{\rm high} \sim -11.45\times 10^{-15}\ \rm s^{-2}$ and State II: $|\dot{\nu}|_{\rm low} \sim -9.76\times 10^{-15}\ \rm s^{-2}$). The ratio of duration in the State I and State II is $\sim 60\%$ and $\sim 20\%$, respectively. The rest of the time pulse profile changes are spotted with complex transitions in the spin-down rate. \cite{bkb14} linked the unexpected presence of a new profile component to the drastic change in its spin-down rate beginning in 2005. One decade after the first case, this pulsar repeated the spin-down rate state change with integrated profile shape variations. Nonetheless, these two mode-switching events display some distinguishing features. The former shows the crucial change on the leading edge whereas the latter presents the leading component recession, accompanied by an obvious enhancement of the middle part of the pulse profile. The fractional amplitude of the spin-down rate change in 2005 event is more than twice that of the recently observed one. Moreover, the peculiar emission properties in 2015 event are associated with a glitch.

There are at least two possible reasons, one internal and one external origin, which can be responsible for the transition from high spin-down rate State I to the low spin-down rate State II for PSR J0738$-$4042 around 2005 September. According to the internal scenario, a thermal runaway will lead to an increase in the radial vortex creep rate. Standard cooling calculations \citep{yakovlev04,page06} predict that a 1 Myr-old neutron star should have a surface temperature well below $10^{5}$ K if no heating processes are present. However, vortex creep against lattice nuclei in the inner crust \citep{alpar84} and magnetic flux tubes in the outer core \citep{erbil17a} give rise to superfluid friction with normal matter which dissipates rotational energy and heats up the neutron star interior. The resulting interior temperature rise will change the coupling between the interior superfluid components and the observed crust which leads to a thermo-rotational instability \citep{greenstein79} by which a substantial decrease in the magnitude of the spin-down rate may occur. The resulting excess creep rate may also unpin vortex lines in the outer core as angular velocity lag here is closer to the critical threshold for catastrophic unpinning conditions \citep{erbil16}. This may also account for the 2007 timing event. As for the external origin case, a debris disk and asteroid belt may form around neutron stars from supernova fallback matter. The tidal disruption and evaporation of asteroid matter by pulsar gravitational field and irradiation, and their interaction with the magnetosphere may also lead spin-down rate transitions. \citet{bkb14} considered the effect of injected charged density on the activation of pair production regions in the magnetospheric gap regions. The pollution of accelerator gaps reduces the braking torque on the pulsar and in turn gives rise to a low spin-down rate. \citet{huang16}, on the other hand, have taken angular momentum transfer from the debris disk onto the surface of the neutron star into account as the main cause of reduction in the spin-down rate of PSR J0738$-$4042.

Identifying new concurrent variations in pulse profile and timing behaviours makes PSR J0738$-$4042
the seventh glitch-driven state-switching pulsar. Of the mode-switching pulsars in Table \ref{tab:Emission}, the rotational and other properties of PSR B2021+51 are most comparable with PSR J0738$-$4042, especially the characteristic age. \cite{lwy21} revealed that the presence of emission state switching phenomenon in PSR B2021+51 manifests itself in wider mean pulse profiles following a small glitch with $\Delta\nu/\nu = 0.373(5)\times 10^{-9}$. In addition, the post-glitch spin-down rate first gradually increased and after a definite time it reversed and started to decay. These features are just like the case of PSR J0738$-$4042. \cite{lwy21} argued that the changes of the integrated profile are a manifestation of the movements of flux tubes in the emission zone during the glitch. According to the scenario presented in this paper, the spin-down rate shown in the bottom panel of Figure 2 of \citet{lwy21} can be fully explainable in terms of the combined vortex creep response of the inward and outward moving vortices during glitch. The fact that its $W_{50}$ pulse width experienced a increment after the glitch is a consequence of the spin-down rate induced quake instead. Another remarkable case discovered in PSR B2035+36 has been explained as a shift in the magnetic inclination angle $\alpha$ due to the occurrence of a glitch \citep{kyw18}. Meanwhile, \cite{kyw18} also speculated that glitch may change the out-flowing particle density in the magnetosphere, leading to a persistent increasing spin-down rate. Although all previous studies considered that these state-switching behaviours are induced by the glitch event, no in-depth analysis based on the mechanism behind glitch has been performed, to interpret these simultaneous variations in profile shape and $\dot{\nu}$. In the superfluid picture, the persistent increase of the spin-down rate following the glitch of PSR B2035+36 can be elucidated in terms of the formation of a new vortex trap at the time of the glitch wherein the vortex current has ceased.


\section{CONCLUSIONS}{\label{conclusions}}

In this work we present the identification of a repeated emission-rotation correlation occurred in PSR J0738$-$4042, by analysing the combined UTMOST and Parkes data. The newly discovered glitch followed by a sustained increase in the spin-down rate over 1380 d is considered to correlate with variations in the shape of the integrated pulse profile. This event is very important to establish interrelation between the magnetospheric processes and the neutron-star internal dynamics. We evaluate the 2015 pulse profile change-spin-down rate variation correlation as being a consequence of the crustquake induced inclination change and vortex line movement. The irreversible displacement of the broken crustal platelet in the direction of the neutron star equator has brought about a tiny increment in the pulsar obliquity which resulted in the narrowing of the pulse profile as is evident from comparison of $W_{50}$ pulse width values before and after of the event. Shift of the platelet also carried vortices attached on it. The associated change of the vortex line configuration was resulted in gradual increase of the spin-down rate which survived on very long time-scale. This scenario enabled us to infer neutron star parameters like the moment of the inertia and recoupling time-scale of the crustal superfluid. Our results given in Table \ref{model} means that PSR J0738$-$4042 should have a thick crust and in turn stiff equation of state. We estimated the total number of vortices that involved in the glitch event. The number of vortices is thousand times smaller than inferred for the large glitches experienced by the Vela-like pulsars. We conclude that for PSR J0738$-$4042 either the typical distance between the vortex traps are large and has an unconnected vortex network or the low temperature of its crust reduces the vortex-vortex scattering rate which is responsible for large-scale unpinning avalanche. We put a lower limit to the surface temperature of this pulsar by invoking superfluid friction with normal matter. We also estimated the size of the broken platelet which seems to be a global pulsar quantity reflecting the magnitude of the critical strain angle of neutron star crystal. 

The model we propose has a predictive feature: if the spin-down rate variation is a consequence of the combined inward and outward motion vortex lines during a crust breaking quake, it will first display a gradual increase and start to decay after a certain time determined by the superfluid recoupling time-scale affected by the glitch. The pulse profile would become wider (narrower) if the underlying physical mechanism for broken platelet motion during the glitch was a crustquake arising from spin-down (vortex line-flux tube pinning) stresses. The ongoing monitoring of PSR J0738$-$4042 at multiple radio wavelengths will certainly be a pragmatic approach to investigate the relationships between the concurrent changes in the pulse shape and timing behaviours, and help to discriminate among different models proposed for mode-switching behaviours discovered in 2005.


\section*{Acknowledgements}
The Parkes radio telescope is part of the Australia Telescope, which is funded by the Commonwealth of Australia for operation as a National Facility managed by the Commonwealth Scientific and Industrial Research Organisation (CSIRO). The Molonglo Observatory is owned and operated by the University of Sydney. Major support for the UTMOST project has been provided by Swinburne University of Technology. Our work is funded by the National Natural Science Foundation of China via NSFC-11373064, 11521303, 11733010, 11873103, U2031121, 11873080, 12105231, U1838201, U1838202, U1838104, U1938103, U1938109, U1731111, U1938117, 11988101, U1731238, and 11703003. S.Q.Z. is also supported by the Sichuan Youth Science and Technology Innovation Research Team (No. 21CXTD0038) and the Fundamental Research Funds of China West Normal University (No. 20B009). Z.W.F. acknowledges support from the Guiding Local Science and Technology Development Projects by the Central Government  of China no. 2021ZYD0031. We are thankful to Professor Simon Johnston for kindly sharing the data in \citet{bkb14} with us. We would like to express our gratitude to everyone who contributed to make this study possible. We appreciate the referee for insightful comments which led to clarify the presentation.

\section*{Data Availability}

The data used to produce figures in this article will be shared upon request to the authors.

\bibliographystyle{mnras}
\bibliography{ShiqiZhou}

\begin{thebibliography}{}
\makeatletter
\relax
\def\mn@urlcharsother{\let\do\@makeother \do\$\do\&\do\#\do\^\do\_\do\%\do\~}
\def\mn@doi{\begingroup\mn@urlcharsother \@ifnextchar [ {\mn@doi@}
  {\mn@doi@[]}}
\def\mn@doi@[#1]#2{\def\@tempa{#1}\ifx\@tempa\@empty \href
  {http://dx.doi.org/#2} {doi:#2}\else \href {http://dx.doi.org/#2} {#1}\fi
  \endgroup}
\def\mn@eprint#1#2{\mn@eprint@#1:#2::\@nil}
\def\mn@eprint@arXiv#1{\href {http://arxiv.org/abs/#1} {{\tt arXiv:#1}}}
\def\mn@eprint@dblp#1{\href {http://dblp.uni-trier.de/rec/bibtex/#1.xml}
  {dblp:#1}}
\def\mn@eprint@#1:#2:#3:#4\@nil{\def\@tempa {#1}\def\@tempb {#2}\def\@tempc
  {#3}\ifx \@tempc \@empty \let \@tempc \@tempb \let \@tempb \@tempa \fi \ifx
  \@tempb \@empty \def\@tempb {arXiv}\fi \@ifundefined
  {mn@eprint@\@tempb}{\@tempb:\@tempc}{\expandafter \expandafter \csname
  mn@eprint@\@tempb\endcsname \expandafter{\@tempc}}}

\bibitem[\protect\citeauthoryear{{Akbal}, {G{\"u}gercino{\u{g}}lu},
  {{\c{S}}a{\c{s}}maz Mu{\c{s}}}  \& {Alpar}}{{Akbal} et~al.}{2015}]{akbal15}
{Akbal} O.,  {G{\"u}gercino{\u{g}}lu} E.,  {{\c{S}}a{\c{s}}maz Mu{\c{s}}} S.,
  {Alpar} M.~A.,  2015, \mn@doi [\mnras] {10.1093/mnras/stv322}, \href
  {https://ui.adsabs.harvard.edu/abs/2015MNRAS.449..933A} {449, 933}

\bibitem[\protect\citeauthoryear{{Akbal}, {Alpar}, {Buchner}  \&
  {Pines}}{{Akbal} et~al.}{2017}]{akbal17}
{Akbal} O.,  {Alpar} M.~A.,  {Buchner} S.,   {Pines} D.,  2017, \mn@doi
  [\mnras] {10.1093/mnras/stx1095}, \href
  {https://ui.adsabs.harvard.edu/abs/2017MNRAS.469.4183A} {469, 4183}

\bibitem[\protect\citeauthoryear{{Allafort} et~al.,}{{Allafort}
  et~al.}{2013}]{abb13}
{Allafort} A.,  et~al., 2013, \mn@doi [\apjl] {10.1088/2041-8205/777/1/L2},
  \href {https://ui.adsabs.harvard.edu/abs/2013ApJ...777L...2A} {777, L2}

\bibitem[\protect\citeauthoryear{{Alpar} \& {Baykal}}{{Alpar} \&
  {Baykal}}{2006}]{ab06}
{Alpar} M.~A.,  {Baykal} A.,  2006, \mn@doi [\mnras]
  {10.1111/j.1365-2966.2006.10893.x}, \href
  {https://ui.adsabs.harvard.edu/abs/2006MNRAS.372..489A} {372, 489}

\bibitem[\protect\citeauthoryear{{Alpar}, {Anderson}, {Pines}  \&
  {Shaham}}{{Alpar} et~al.}{1984}]{alpar84}
{Alpar} M.~A.,  {Anderson} P.~W.,  {Pines} D.,   {Shaham} J.,  1984, \mn@doi
  [\apj] {10.1086/161616}, \href
  {https://ui.adsabs.harvard.edu/abs/1984ApJ...276..325A} {276, 325}

\bibitem[\protect\citeauthoryear{{Alpar}, {Chau}, {Cheng}  \& {Pines}}{{Alpar}
  et~al.}{1996}]{alpar96}
{Alpar} M.~A.,  {Chau} H.~F.,  {Cheng} K.~S.,   {Pines} D.,  1996, \mn@doi
  [\apj] {10.1086/176935}, \href
  {https://ui.adsabs.harvard.edu/abs/1996ApJ...459..706A} {459, 706}

\bibitem[\protect\citeauthoryear{{Anderson}, {Alpar}, {Pines}  \&
  {Shaham}}{{Anderson} et~al.}{1982}]{anderson82}
{Anderson} P.~W.,  {Alpar} M.~A.,  {Pines} D.,   {Shaham} J.,  1982, \mn@doi
  [Philosophical Magazine, Part A] {10.1080/01418618208244296}, \href
  {https://ui.adsabs.harvard.edu/abs/1982PMagA..45..227A} {45, 227}

\bibitem[\protect\citeauthoryear{{Arzoumanian}, {Nice}, {Taylor}  \&
  {Thorsett}}{{Arzoumanian} et~al.}{1994}]{arzoumanian94}
{Arzoumanian} Z.,  {Nice} D.~J.,  {Taylor} J.~H.,   {Thorsett} S.~E.,  1994,
  \mn@doi [\apj] {10.1086/173760}, \href
  {https://ui.adsabs.harvard.edu/abs/1994ApJ...422..671A} {422, 671}

\bibitem[\protect\citeauthoryear{{Ashton}, {Lasky}, {Graber}  \&
  {Palfreyman}}{{Ashton} et~al.}{2019}]{alg19}
{Ashton} G.,  {Lasky} P.~D.,  {Graber} V.,   {Palfreyman} J.,  2019, \mn@doi
  [Nature Astronomy] {10.1038/s41550-019-0844-6}, \href
  {https://ui.adsabs.harvard.edu/abs/2019NatAs...3.1143A} {3, 1143}

\bibitem[\protect\citeauthoryear{{Basu} et~al.,}{{Basu} et~al.}{2022}]{bsa22}
{Basu} A.,  et~al., 2022, \mn@doi [\mnras] {10.1093/mnras/stab3336}, \href
  {https://ui.adsabs.harvard.edu/abs/2022MNRAS.510.4049B} {510, 4049}

\bibitem[\protect\citeauthoryear{{Baym} \& {Pines}}{{Baym} \&
  {Pines}}{1971}]{baym71}
{Baym} G.,  {Pines} D.,  1971, \mn@doi [Annals of Physics]
  {10.1016/0003-4916(71)90084-4}, \href
  {https://ui.adsabs.harvard.edu/abs/1971AnPhy..66..816B} {66, 816}

\bibitem[\protect\citeauthoryear{{Boynton}, {Groth}, {Hutchinson}, {Nanos},
  {Partridge}  \& {Wilkinson}}{{Boynton} et~al.}{1972}]{bgh72}
{Boynton} P.~E.,  {Groth} E.~J.,  {Hutchinson} D.~P.,  {Nanos} G.~P. J.,
  {Partridge} R.~B.,   {Wilkinson} D.~T.,  1972, \mn@doi [\apj]
  {10.1086/151550}, \href
  {https://ui.adsabs.harvard.edu/abs/1972ApJ...175..217B} {175, 217}

\bibitem[\protect\citeauthoryear{{Brook}, {Karastergiou}, {Buchner}, {Roberts},
  {Keith}, {Johnston}  \& {Shannon}}{{Brook} et~al.}{2014}]{bkb14}
{Brook} P.~R.,  {Karastergiou} A.,  {Buchner} S.,  {Roberts} S.~J.,  {Keith}
  M.~J.,  {Johnston} S.,   {Shannon} R.~M.,  2014, \mn@doi [\apjl]
  {10.1088/2041-8205/780/2/L31}, \href
  {https://ui.adsabs.harvard.edu/abs/2014ApJ...780L..31B} {780, L31}

\bibitem[\protect\citeauthoryear{{Brook}, {Karastergiou}, {Johnston}, {Kerr},
  {Shannon}  \& {Roberts}}{{Brook} et~al.}{2016}]{bkj16}
{Brook} P.~R.,  {Karastergiou} A.,  {Johnston} S.,  {Kerr} M.,  {Shannon}
  R.~M.,   {Roberts} S.~J.,  2016, \mnras, \href
  {http://adsabs.harvard.edu/abs/2016MNRAS.456.1374B} {456, 1374}

\bibitem[\protect\citeauthoryear{{Cheng}}{{Cheng}}{1987}]{cheng87}
{Cheng} K.~S.,  1987, \mn@doi [\apj] {10.1086/165673}, \href
  {https://ui.adsabs.harvard.edu/abs/1987ApJ...321..805C} {321, 805}

\bibitem[\protect\citeauthoryear{{Cheng}, {Alpar}, {Pines}  \&
  {Shaham}}{{Cheng} et~al.}{1988}]{cheng88}
{Cheng} K.~S.,  {Alpar} M.~A.,  {Pines} D.,   {Shaham} J.,  1988, \mn@doi
  [\apj] {10.1086/166517}, \href
  {https://ui.adsabs.harvard.edu/abs/1988ApJ...330..835C} {330, 835}

\bibitem[\protect\citeauthoryear{{Chukwude} \& {Buchner}}{{Chukwude} \&
  {Buchner}}{2012}]{cb12}
{Chukwude} A.~E.,  {Buchner} S.,  2012, \mn@doi [\apj]
  {10.1088/0004-637X/745/1/40}, \href
  {https://ui.adsabs.harvard.edu/abs/2012ApJ...745...40C} {745, 40}

\bibitem[\protect\citeauthoryear{{Cordes} \& {Downs}}{{Cordes} \&
  {Downs}}{1985}]{Cordes1985}
{Cordes} J.~M.,  {Downs} G.~S.,  1985, \mn@doi [\apjs] {10.1086/191076}, \href
  {https://ui.adsabs.harvard.edu/abs/1985ApJS...59..343C} {59, 343}

\bibitem[\protect\citeauthoryear{{Dang} et~al.,}{{Dang} et~al.}{2020}]{dym20}
{Dang} S.~J.,  et~al., 2020, \mn@doi [\apj] {10.3847/1538-4357/ab9082}, \href
  {https://ui.adsabs.harvard.edu/abs/2020ApJ...896..140D} {896, 140}

\bibitem[\protect\citeauthoryear{{Dang} et~al.,}{{Dang} et~al.}{2021}]{dww21}
{Dang} S.~J.,  et~al., 2021, \mn@doi [RAA] {10.1088/1674-4527/21/2/42}, \href
  {https://ui.adsabs.harvard.edu/abs/2021RAA....21...42D} {21, 042}

\bibitem[\protect\citeauthoryear{{Delsate}, {Chamel}, {G{\"u}rlebeck},
  {Fantina}, {Pearson}  \& {Ducoin}}{{Delsate} et~al.}{2016}]{delsate16}
{Delsate} T.,  {Chamel} N.,  {G{\"u}rlebeck} N.,  {Fantina} A.~F.,  {Pearson}
  J.~M.,   {Ducoin} C.,  2016, \mn@doi [\prd] {10.1103/PhysRevD.94.023008},
  \href {https://ui.adsabs.harvard.edu/abs/2016PhRvD..94b3008D} {94, 023008}

\bibitem[\protect\citeauthoryear{{Dib} \& {Kaspi}}{{Dib} \&
  {Kaspi}}{2014}]{dk14}
{Dib} R.,  {Kaspi} V.~M.,  2014, \apj, \href
  {http://adsabs.harvard.edu/abs/2014ApJ...784...37D} {784, 37}

\bibitem[\protect\citeauthoryear{{Douchin} \& {Haensel}}{{Douchin} \&
  {Haensel}}{2001}]{douchin01}
{Douchin} F.,  {Haensel} P.,  2001, \mn@doi [\aap]
  {10.1051/0004-6361:20011402}, \href
  {https://ui.adsabs.harvard.edu/abs/2001A&A...380..151D} {380, 151}

\bibitem[\protect\citeauthoryear{{Downs} \& {Krause-Polstorff}}{{Downs} \&
  {Krause-Polstorff}}{1986}]{Downs1986}
{Downs} G.~S.,  {Krause-Polstorff} J.,  1986, \mn@doi [\apjs] {10.1086/191134},
  \href {https://ui.adsabs.harvard.edu/abs/1986ApJS...62...81D} {62, 81}

\bibitem[\protect\citeauthoryear{{Downs} \& {Reichley}}{{Downs} \&
  {Reichley}}{1983}]{Downs1983}
{Downs} G.~S.,  {Reichley} P.~E.,  1983, \mn@doi [\apjs] {10.1086/190890},
  \href {https://ui.adsabs.harvard.edu/abs/1983ApJS...53..169D} {53, 169}

\bibitem[\protect\citeauthoryear{{Dunn} et~al.,}{{Dunn} et~al.}{2022}]{dms22}
{Dunn} L.,  et~al., 2022, \mn@doi [\mnras] {10.1093/mnras/stac551}, \href
  {https://ui.adsabs.harvard.edu/abs/2022MNRAS.512.1469D} {512, 1469}

\bibitem[\protect\citeauthoryear{{Edwards}, {Hobbs}  \& {Manchester}}{{Edwards}
  et~al.}{2006}]{ehm06}
{Edwards} R.~T.,  {Hobbs} G.~B.,   {Manchester} R.~N.,  2006, \mnras, \href
  {http://adsabs.harvard.edu/abs/2006MNRAS.372.1549E} {372, 1549}

\bibitem[\protect\citeauthoryear{{Ek{\c{s}}i}, {Anda{\c{c}}},
  {{\c{C}}{\i}k{\i}nto{\u{g}}lu}, {G{\"u}gercino{\u{g}}lu}, {Vahdat Motlagh}
  \& {K{\i}z{\i}ltan}}{{Ek{\c{s}}i} et~al.}{2016}]{eksi16}
{Ek{\c{s}}i} K.~Y.,  {Anda{\c{c}}} I.~C.,  {{\c{C}}{\i}k{\i}nto{\u{g}}lu} S.,
  {G{\"u}gercino{\u{g}}lu} E.,  {Vahdat Motlagh} A.,   {K{\i}z{\i}ltan} B.,
  2016, \mn@doi [\apj] {10.3847/0004-637X/823/1/34}, \href
  {https://ui.adsabs.harvard.edu/abs/2016ApJ...823...34E} {823, 34}

\bibitem[\protect\citeauthoryear{{Espinoza}, {Lyne}, {Stappers}  \&
  {Kramer}}{{Espinoza} et~al.}{2011}]{espinoza11}
{Espinoza} C.~M.,  {Lyne} A.~G.,  {Stappers} B.~W.,   {Kramer} M.,  2011,
  \mn@doi [\mnras] {10.1111/j.1365-2966.2011.18503.x}, \href
  {https://ui.adsabs.harvard.edu/abs/2011MNRAS.414.1679E} {414, 1679}

\bibitem[\protect\citeauthoryear{{Folkner}, {Williams}, {Boggs}, {Park}  \&
  {Kuchynka}}{{Folkner} et~al.}{2014}]{ftb14}
{Folkner} W.~M.,  {Williams} J.~G.,  {Boggs} D.~H.,  {Park} R.~S.,   {Kuchynka}
  P.,  2014, Interplanetary Network Progress Report, \href
  {https://ui.adsabs.harvard.edu/abs/2014IPNPR.196C...1F} {42-196, 1}

\bibitem[\protect\citeauthoryear{{Franco}, {Link}  \& {Epstein}}{{Franco}
  et~al.}{2000}]{franco00}
{Franco} L.~M.,  {Link} B.,   {Epstein} R.~I.,  2000, \mn@doi [\apj]
  {10.1086/317121}, \href
  {https://ui.adsabs.harvard.edu/abs/2000ApJ...543..987F} {543, 987}

\bibitem[\protect\citeauthoryear{{Ge} et~al.,}{{Ge} et~al.}{2020a}]{ge20}
{Ge} M.~Y.,  et~al., 2020a, \mn@doi [\apj] {10.3847/1538-4357/ab8db6}, \href
  {https://ui.adsabs.harvard.edu/abs/2020ApJ...896...55G} {896, 55}

\bibitem[\protect\citeauthoryear{{Ge} et~al.,}{{Ge} et~al.}{2020b}]{ge20a}
{Ge} M.~Y.,  et~al., 2020b, \mn@doi [\apjl] {10.3847/2041-8213/abaeed}, \href
  {https://ui.adsabs.harvard.edu/abs/2020ApJ...900L...7G} {900, L7}

\bibitem[\protect\citeauthoryear{{Geppert}, {Basu}, {Mitra}, {Melikidze}  \&
  {Szkudlarek}}{{Geppert} et~al.}{2021}]{geppert21}
{Geppert} U.,  {Basu} R.,  {Mitra} D.,  {Melikidze} G.~I.,   {Szkudlarek} M.,
  2021, \mn@doi [\mnras] {10.1093/mnras/stab1134}, \href
  {https://ui.adsabs.harvard.edu/abs/2021MNRAS.504.5741G} {504, 5741}

\bibitem[\protect\citeauthoryear{{Giliberti}, {Cambiotti}, {Antonelli}  \&
  {Pizzochero}}{{Giliberti} et~al.}{2020}]{giliberti20}
{Giliberti} E.,  {Cambiotti} G.,  {Antonelli} M.,   {Pizzochero} P.~M.,  2020,
  \mn@doi [\mnras] {10.1093/mnras/stz3099}, \href
  {https://ui.adsabs.harvard.edu/abs/2020MNRAS.491.1064G} {491, 1064}

\bibitem[\protect\citeauthoryear{{Gonz{\'a}lez-Jim{\'e}nez}, {Petrovich}  \&
  {Reisenegger}}{{Gonz{\'a}lez-Jim{\'e}nez} et~al.}{2015}]{reisenegger15}
{Gonz{\'a}lez-Jim{\'e}nez} N.,  {Petrovich} C.,   {Reisenegger} A.,  2015,
  \mn@doi [\mnras] {10.1093/mnras/stu2558}, \href
  {https://ui.adsabs.harvard.edu/abs/2015MNRAS.447.2073G} {447, 2073}

\bibitem[\protect\citeauthoryear{{Greenstein}}{{Greenstein}}{1979}]{greenstein79}
{Greenstein} G.,  1979, \mn@doi [\nat] {10.1038/277521a0}, \href
  {https://ui.adsabs.harvard.edu/abs/1979Natur.277..521G} {277, 521}

\bibitem[\protect\citeauthoryear{{Gudmundsson}, {Pethick}  \&
  {Epstein}}{{Gudmundsson} et~al.}{1983}]{gudmundsson83}
{Gudmundsson} E.~H.,  {Pethick} C.~J.,   {Epstein} R.~I.,  1983, \mn@doi [\apj]
  {10.1086/161292}, \href
  {https://ui.adsabs.harvard.edu/abs/1983ApJ...272..286G} {272, 286}

\bibitem[\protect\citeauthoryear{{G{\"u}gercino{\u{g}}lu}}{{G{\"u}gercino{\u{g}}lu}}{2017}]{erbil17a}
{G{\"u}gercino{\u{g}}lu} E.,  2017, \mn@doi [\mnras] {10.1093/mnras/stx985},
  \href {https://ui.adsabs.harvard.edu/abs/2017MNRAS.469.2313G} {469, 2313}

\bibitem[\protect\citeauthoryear{{G{\"u}gercino{\u{g}}lu} \&
  {Alpar}}{{G{\"u}gercino{\u{g}}lu} \& {Alpar}}{2014}]{erbil14}
{G{\"u}gercino{\u{g}}lu} E.,  {Alpar} M.~A.,  2014, \mn@doi [\apjl]
  {10.1088/2041-8205/788/1/L11}, \href
  {https://ui.adsabs.harvard.edu/abs/2014ApJ...788L..11G} {788, L11}

\bibitem[\protect\citeauthoryear{{G{\"u}gercino{\u{g}}lu} \&
  {Alpar}}{{G{\"u}gercino{\u{g}}lu} \& {Alpar}}{2016}]{erbil16}
{G{\"u}gercino{\u{g}}lu} E.,  {Alpar} M.~A.,  2016, \mn@doi [\mnras]
  {10.1093/mnras/stw1758}, \href
  {https://ui.adsabs.harvard.edu/abs/2016MNRAS.462.1453G} {462, 1453}

\bibitem[\protect\citeauthoryear{{G{\"u}gercino{\u{g}}lu} \&
  {Alpar}}{{G{\"u}gercino{\u{g}}lu} \& {Alpar}}{2020}]{erbil20}
{G{\"u}gercino{\u{g}}lu} E.,  {Alpar} M.~A.,  2020, \mn@doi [\mnras]
  {10.1093/mnras/staa1672}, \href
  {https://ui.adsabs.harvard.edu/abs/2020MNRAS.496.2506G} {496, 2506}

\bibitem[\protect\citeauthoryear{{G{\"u}gercino{\u{g}}lu}, {K{\"o}ksal}  \&
  {G{\"u}ver}}{{G{\"u}gercino{\u{g}}lu} et~al.}{2022a}]{erbil2022}
{G{\"u}gercino{\u{g}}lu} E.,  {K{\"o}ksal} E.,   {G{\"u}ver} T.,  2022a, arXiv
  e-prints, \href {https://ui.adsabs.harvard.edu/abs/2022arXiv220704111G} {p.
  arXiv:2207.04111}

\bibitem[\protect\citeauthoryear{{G{\"u}gercino{\u{g}}lu}, {Ge}, {Yuan}  \&
  {Zhou}}{{G{\"u}gercino{\u{g}}lu} et~al.}{2022b}]{egy22}
{G{\"u}gercino{\u{g}}lu} E.,  {Ge} M.~Y.,  {Yuan} J.~P.,   {Zhou} S.~Q.,
  2022b, \mn@doi [\mnras] {10.1093/mnras/stac026}, \href
  {https://ui.adsabs.harvard.edu/abs/2022MNRAS.511..425G} {511, 425}

\bibitem[\protect\citeauthoryear{{G{\"u}gercino{\v{g}}lu} \&
  {Alpar}}{{G{\"u}gercino{\v{g}}lu} \& {Alpar}}{2017}]{erbil17}
{G{\"u}gercino{\v{g}}lu} E.,  {Alpar} M.~A.,  2017, \mn@doi [\mnras]
  {10.1093/mnras/stx1937}, \href
  {https://ui.adsabs.harvard.edu/abs/2017MNRAS.471.4827G} {471, 4827}

\bibitem[\protect\citeauthoryear{{G{\"u}gercino{\v{g}}lu} \&
  {Alpar}}{{G{\"u}gercino{\v{g}}lu} \& {Alpar}}{2019}]{erbil19}
{G{\"u}gercino{\v{g}}lu} E.,  {Alpar} M.~A.,  2019, \mn@doi [\mnras]
  {10.1093/mnras/stz1831}, \href
  {https://ui.adsabs.harvard.edu/abs/2019MNRAS.488.2275G} {488, 2275}

\bibitem[\protect\citeauthoryear{{Ho}, {Espinoza}, {Antonopoulou}  \&
  {Andersson}}{{Ho} et~al.}{2015}]{ho15}
{Ho} W.~C.~G.,  {Espinoza} C.~M.,  {Antonopoulou} D.,   {Andersson} N.,  2015,
  \mn@doi [Science Advances] {10.1126/sciadv.1500578}, \href
  {https://ui.adsabs.harvard.edu/abs/2015SciA....1E0578H} {1, e1500578}

\bibitem[\protect\citeauthoryear{{Hobbs}, {Edwards}  \& {Manchester}}{{Hobbs}
  et~al.}{2006}]{hem06}
{Hobbs} G.~B.,  {Edwards} R.~T.,   {Manchester} R.~N.,  2006, \mnras, \href
  {http://adsabs.harvard.edu/abs/2006MNRAS.369..655H} {369, 655}

\bibitem[\protect\citeauthoryear{{Hobbs}, {Lyne}  \& {Kramer}}{{Hobbs}
  et~al.}{2010}]{hlk10}
{Hobbs} G.,  {Lyne} A.~G.,   {Kramer} M.,  2010, \mn@doi [\mnras]
  {10.1111/j.1365-2966.2009.15938.x}, \href
  {https://ui.adsabs.harvard.edu/abs/2010MNRAS.402.1027H} {402, 1027}

\bibitem[\protect\citeauthoryear{{Hobbs} et~al.,}{{Hobbs} et~al.}{2011}]{hmm11}
{Hobbs} G.,  et~al., 2011, \pasa, \href
  {http://adsabs.harvard.edu/abs/2011PASA...28..202H} {28, 202}

\bibitem[\protect\citeauthoryear{{Hotan}, {van Straten}  \&
  {Manchester}}{{Hotan} et~al.}{2004}]{hvm04}
{Hotan} A.~W.,  {van Straten} W.,   {Manchester} R.~N.,  2004, \pasa, \href
  {http://adsabs.harvard.edu/abs/2004PASA...21..302H} {21, 302}

\bibitem[\protect\citeauthoryear{{Jankowski} et~al.,}{{Jankowski}
  et~al.}{2019}]{jbs19}
{Jankowski} F.,  et~al., 2019, \mn@doi [\mnras] {10.1093/mnras/sty3390}, \href
  {https://ui.adsabs.harvard.edu/abs/2019MNRAS.484.3691J} {484, 3691}

\bibitem[\protect\citeauthoryear{{Johnston} \& {Galloway}}{{Johnston} \&
  {Galloway}}{1999}]{jg99}
{Johnston} S.,  {Galloway} D.,  1999, \mn@doi [\mnras]
  {10.1046/j.1365-8711.1999.02737.x}, \href
  {https://ui.adsabs.harvard.edu/abs/1999MNRAS.306L..50J} {306, L50}

\bibitem[\protect\citeauthoryear{{Jones}}{{Jones}}{1990}]{jones90}
{Jones} P.~B.,  1990, \mnras, \href
  {https://ui.adsabs.harvard.edu/abs/1990MNRAS.246..364J} {246, 364}

\bibitem[\protect\citeauthoryear{{Karastergiou}, {Roberts}, {Johnston}, {Lee},
  {Weltevrede}  \& {Kramer}}{{Karastergiou} et~al.}{2011}]{krj11}
{Karastergiou} A.,  {Roberts} S.~J.,  {Johnston} S.,  {Lee} H.,  {Weltevrede}
  P.,   {Kramer} M.,  2011, \mn@doi [\mnras]
  {10.1111/j.1365-2966.2011.18697.x}, \href
  {https://ui.adsabs.harvard.edu/abs/2011MNRAS.415..251K} {415, 251}

\bibitem[\protect\citeauthoryear{{Keith}, {Shannon}  \& {Johnston}}{{Keith}
  et~al.}{2013}]{ksj13}
{Keith} M.~J.,  {Shannon} R.~M.,   {Johnston} S.,  2013, \mnras, \href
  {http://adsabs.harvard.edu/abs/2013MNRAS.432.3080K} {432, 3080}

\bibitem[\protect\citeauthoryear{{Kerr}, {Hobbs}, {Johnston}  \&
  {Shannon}}{{Kerr} et~al.}{2016}]{kerr16}
{Kerr} M.,  {Hobbs} G.,  {Johnston} S.,   {Shannon} R.~M.,  2016, \mn@doi
  [\mnras] {10.1093/mnras/stv2457}, \href
  {https://ui.adsabs.harvard.edu/abs/2016MNRAS.455.1845K} {455, 1845}

\bibitem[\protect\citeauthoryear{{Kou}, {Yuan}, {Wang}, {Yan}  \& {Dang}}{{Kou}
  et~al.}{2018}]{kyw18}
{Kou} F.~F.,  {Yuan} J.~P.,  {Wang} N.,  {Yan} W.~M.,   {Dang} S.~J.,  2018,
  \mnras, \href {http://adsabs.harvard.edu/abs/2018MNRAS.478L..24K} {478, L24}

\bibitem[\protect\citeauthoryear{{Lander}, {Andersson}, {Antonopoulou}  \&
  {Watts}}{{Lander} et~al.}{2015}]{lander15}
{Lander} S.~K.,  {Andersson} N.,  {Antonopoulou} D.,   {Watts} A.~L.,  2015,
  \mn@doi [\mnras] {10.1093/mnras/stv432}, \href
  {https://ui.adsabs.harvard.edu/abs/2015MNRAS.449.2047L} {449, 2047}

\bibitem[\protect\citeauthoryear{{Large}, {Vaughan}  \& {Wielebinski}}{{Large}
  et~al.}{1968}]{lvw68}
{Large} M.~I.,  {Vaughan} A.~E.,   {Wielebinski} R.,  1968, \mn@doi [\nat]
  {10.1038/220753a0}, \href
  {https://ui.adsabs.harvard.edu/abs/1968Natur.220..753L} {220, 753}

\bibitem[\protect\citeauthoryear{{Link} \& {Epstein}}{{Link} \&
  {Epstein}}{1996}]{link96}
{Link} B.,  {Epstein} R.~I.,  1996, \mn@doi [\apj] {10.1086/176779}, \href
  {https://ui.adsabs.harvard.edu/abs/1996ApJ...457..844L} {457, 844}

\bibitem[\protect\citeauthoryear{{Link} \& {Epstein}}{{Link} \&
  {Epstein}}{1997}]{le97}
{Link} B.,  {Epstein} R.~I.,  1997, \mn@doi [\apjl] {10.1086/310549}, \href
  {https://ui.adsabs.harvard.edu/abs/1997ApJ...478L..91L} {478, L91}

\bibitem[\protect\citeauthoryear{{Liu}, {Zhou}, {Zhang}, {Feng}  \&
  {Zhou}}{{Liu} et~al.}{2021a}]{liu21}
{Liu} H.-Y.,  {Zhou} S.-Q.,  {Zhang} Y.-Q.,  {Feng} Z.-W.,   {Zhou} X.,  2021a,
  \mn@doi [RAA] {10.1088/1674-4527/21/7/154}, \href
  {https://ui.adsabs.harvard.edu/abs/2021RAA....21..154L} {21, 154}

\bibitem[\protect\citeauthoryear{{Liu}, {Wang}, {Yan}, {Shen}, {Tong}, {Huang}
  \& {Zhao}}{{Liu} et~al.}{2021b}]{lwy21}
{Liu} J.,  {Wang} H.-G.,  {Yan} Z.,  {Shen} Z.-Q.,  {Tong} H.,  {Huang} Z.-P.,
   {Zhao} R.-S.,  2021b, \mn@doi [\apj] {10.3847/1538-4357/abf140}, \href
  {https://ui.adsabs.harvard.edu/abs/2021ApJ...912...58L} {912, 58}

\bibitem[\protect\citeauthoryear{{Liu}, {Wang}, {Shen}, {Yan}, {Tong}, {Huang}
  \& {Zhao}}{{Liu} et~al.}{2022}]{lws22}
{Liu} J.,  {Wang} H.-G.,  {Shen} Z.-Q.,  {Yan} Z.,  {Tong} H.,  {Huang} Z.-P.,
   {Zhao} R.-S.,  2022, \mn@doi [\apj] {10.3847/1538-4357/ac6bf7}, \href
  {https://ui.adsabs.harvard.edu/abs/2022ApJ...931..103L} {931, 103}

\bibitem[\protect\citeauthoryear{{Lower} et~al.,}{{Lower} et~al.}{2020}]{lbs20}
{Lower} M.~E.,  et~al., 2020, \mn@doi [\mnras] {10.1093/mnras/staa615}, \href
  {https://ui.adsabs.harvard.edu/abs/2020MNRAS.494..228L} {494, 228}

\bibitem[\protect\citeauthoryear{{Lower} et~al.,}{{Lower} et~al.}{2021}]{ljd21}
{Lower} M.~E.,  et~al., 2021, \mn@doi [\mnras] {10.1093/mnras/stab2678}, \href
  {https://ui.adsabs.harvard.edu/abs/2021MNRAS.508.3251L} {508, 3251}

\bibitem[\protect\citeauthoryear{{Lyne}, {Hobbs}, {Kramer}, {Stairs}  \&
  {Stappers}}{{Lyne} et~al.}{2010}]{lhk10}
{Lyne} A.,  {Hobbs} G.,  {Kramer} M.,  {Stairs} I.,   {Stappers} B.,  2010,
  \mn@doi [Science] {10.1126/science.1186683}, \href
  {https://ui.adsabs.harvard.edu/abs/2010Sci...329..408L} {329, 408}

\bibitem[\protect\citeauthoryear{{Maciesiak} \& {Gil}}{{Maciesiak} \&
  {Gil}}{2011}]{gil11}
{Maciesiak} K.,  {Gil} J.,  2011, \mn@doi [\mnras]
  {10.1111/j.1365-2966.2011.19359.x}, \href
  {https://ui.adsabs.harvard.edu/abs/2011MNRAS.417.1444M} {417, 1444}

\bibitem[\protect\citeauthoryear{{Manchester}}{{Manchester}}{2017}]{man18}
{Manchester} R.~N.,  2017, Proc. Int. Astron. Union, \href
  {https://ui.adsabs.harvard.edu/abs/2018IAUS..337..197M} {13, 197}

\bibitem[\protect\citeauthoryear{{Manchester} \& {Hobbs}}{{Manchester} \&
  {Hobbs}}{2011}]{mh11}
{Manchester} R.~N.,  {Hobbs} G.,  2011, \apjl, \href
  {http://adsabs.harvard.edu/abs/2011ApJ...736L..31M} {736, L31}

\bibitem[\protect\citeauthoryear{{Manchester}, {Newton}, {Hamilton}  \&
  {Goss}}{{Manchester} et~al.}{1983}]{Manchester1983}
{Manchester} R.~N.,  {Newton} L.~M.,  {Hamilton} P.~A.,   {Goss} W.~M.,  1983,
  \mn@doi [\mnras] {10.1093/mnras/202.2.269}, \href
  {https://ui.adsabs.harvard.edu/abs/1983MNRAS.202..269M} {202, 269}

\bibitem[\protect\citeauthoryear{{Manchester}, {Hobbs}, {Teoh}  \&
  {Hobbs}}{{Manchester} et~al.}{2005}]{manchester05}
{Manchester} R.~N.,  {Hobbs} G.~B.,  {Teoh} A.,   {Hobbs} M.,  2005, \mn@doi
  [\aj] {10.1086/428488}, \href
  {https://ui.adsabs.harvard.edu/abs/2005AJ....129.1993M} {129, 1993}

\bibitem[\protect\citeauthoryear{{McCulloch}, {Hamilton}, {McConnell}  \&
  {King}}{{McCulloch} et~al.}{1990}]{mcculloch90}
{McCulloch} P.~M.,  {Hamilton} P.~A.,  {McConnell} D.,   {King} E.~A.,  1990,
  \mn@doi [\nat] {10.1038/346822a0}, \href
  {https://ui.adsabs.harvard.edu/abs/1990Natur.346..822M} {346, 822}

\bibitem[\protect\citeauthoryear{{McKenna} \& {Lyne}}{{McKenna} \&
  {Lyne}}{1990}]{ml90}
{McKenna} J.,  {Lyne} A.~G.,  1990, \nat, \href
  {http://adsabs.harvard.edu/abs/1990Natur.343..349M} {343, 349}

\bibitem[\protect\citeauthoryear{{Montoli}, {Antonelli}, {Magistrelli}  \&
  {Pizzochero}}{{Montoli} et~al.}{2020}]{montoli20}
{Montoli} A.,  {Antonelli} M.,  {Magistrelli} F.,   {Pizzochero} P.~M.,  2020,
  \mn@doi [\aap] {10.1051/0004-6361/202038340}, \href
  {https://ui.adsabs.harvard.edu/abs/2020A&A...642A.223M} {642, A223}

\bibitem[\protect\citeauthoryear{{Ng}, {Takata}  \& {Cheng}}{{Ng}
  et~al.}{2016}]{ntc16}
{Ng} C.~W.,  {Takata} J.,   {Cheng} K.~S.,  2016, \mn@doi [\apj]
  {10.3847/0004-637X/825/1/18}, \href
  {https://ui.adsabs.harvard.edu/abs/2016ApJ...825...18N} {825, 18}

\bibitem[\protect\citeauthoryear{{Page}, {Geppert}  \& {Weber}}{{Page}
  et~al.}{2006}]{page06}
{Page} D.,  {Geppert} U.,   {Weber} F.,  2006, \mn@doi [\nphysa]
  {10.1016/j.nuclphysa.2005.09.019}, \href
  {https://ui.adsabs.harvard.edu/abs/2006NuPhA.777..497P} {777, 497}

\bibitem[\protect\citeauthoryear{{Palfreyman}, {Dickey}, {Hotan}, {Ellingsen}
  \& {van Straten}}{{Palfreyman} et~al.}{2018}]{pdh18}
{Palfreyman} J.,  {Dickey} J.~M.,  {Hotan} A.,  {Ellingsen} S.,   {van Straten}
  W.,  2018, \mn@doi [\nat] {10.1038/s41586-018-0001-x}, \href
  {http://adsabs.harvard.edu/abs/2018Natur.556..219P} {556, 219}

\bibitem[\protect\citeauthoryear{{Philippov}, {Tchekhovskoy}  \&
  {Li}}{{Philippov} et~al.}{2014}]{philippov14}
{Philippov} A.,  {Tchekhovskoy} A.,   {Li} J.~G.,  2014, \mn@doi [\mnras]
  {10.1093/mnras/stu591}, \href
  {https://ui.adsabs.harvard.edu/abs/2014MNRAS.441.1879P} {441, 1879}

\bibitem[\protect\citeauthoryear{{Posselt} et~al.,}{{Posselt}
  et~al.}{2021}]{posselt21}
{Posselt} B.,  et~al., 2021, \mn@doi [\mnras] {10.1093/mnras/stab2775}, \href
  {https://ui.adsabs.harvard.edu/abs/2021MNRAS.508.4249P} {508, 4249}

\bibitem[\protect\citeauthoryear{{Rankin}}{{Rankin}}{1990}]{rankin90}
{Rankin} J.~M.,  1990, \mn@doi [\apj] {10.1086/168530}, \href
  {https://ui.adsabs.harvard.edu/abs/1990ApJ...352..247R} {352, 247}

\bibitem[\protect\citeauthoryear{{Rencoret}, {Aguilera-G{\'o}mez}  \&
  {Reisenegger}}{{Rencoret} et~al.}{2021}]{reisenegger21}
{Rencoret} J.~A.,  {Aguilera-G{\'o}mez} C.,   {Reisenegger} A.,  2021, \mn@doi
  [\aap] {10.1051/0004-6361/202141499}, \href
  {https://ui.adsabs.harvard.edu/abs/2021A&A...654A..47R} {654, A47}

\bibitem[\protect\citeauthoryear{{Ruderman}}{{Ruderman}}{1976}]{ruderman76}
{Ruderman} M.,  1976, \mn@doi [\apj] {10.1086/154069}, \href
  {https://ui.adsabs.harvard.edu/abs/1976ApJ...203..213R} {203, 213}

\bibitem[\protect\citeauthoryear{{Ruderman}, {Zhu}  \& {Chen}}{{Ruderman}
  et~al.}{1998}]{ruderman98}
{Ruderman} M.,  {Zhu} T.,   {Chen} K.,  1998, \mn@doi [\apj] {10.1086/305026},
  \href {https://ui.adsabs.harvard.edu/abs/1998ApJ...492..267R} {492, 267}

\bibitem[\protect\citeauthoryear{{Sedrakian} \& {Cordes}}{{Sedrakian} \&
  {Cordes}}{1999}]{sedrakian99}
{Sedrakian} A.,  {Cordes} J.~M.,  1999, \mn@doi [\mnras]
  {10.1046/j.1365-8711.1999.02638.x}, \href
  {https://ui.adsabs.harvard.edu/abs/1999MNRAS.307..365S} {307, 365}

\bibitem[\protect\citeauthoryear{{Sedrakian} \& {Sedrakian}}{{Sedrakian} \&
  {Sedrakian}}{1993}]{sedrakian93}
{Sedrakian} A.~D.,  {Sedrakian} D.~M.,  1993, \mn@doi [\apj] {10.1086/173034},
  \href {https://ui.adsabs.harvard.edu/abs/1993ApJ...413..658S} {413, 658}

\bibitem[\protect\citeauthoryear{{Seveso}, {Pizzochero}, {Grill}  \&
  {Haskell}}{{Seveso} et~al.}{2016}]{seveso16}
{Seveso} S.,  {Pizzochero} P.~M.,  {Grill} F.,   {Haskell} B.,  2016, \mn@doi
  [\mnras] {10.1093/mnras/stv2579}, \href
  {https://ui.adsabs.harvard.edu/abs/2016MNRAS.455.3952S} {455, 3952}

\bibitem[\protect\citeauthoryear{{Shabanova}}{{Shabanova}}{2007}]{s07}
{Shabanova} T.~V.,  2007, \mn@doi [\apss] {10.1007/s10509-007-9302-5}, \href
  {https://ui.adsabs.harvard.edu/abs/2007Ap&SS.308..591S} {308, 591}

\bibitem[\protect\citeauthoryear{{Shang} \& {Li}}{{Shang} \& {Li}}{2021}]{li21}
{Shang} X.,  {Li} A.,  2021, \mn@doi [\apj] {10.3847/1538-4357/ac2e94}, \href
  {https://ui.adsabs.harvard.edu/abs/2021ApJ...923..108S} {923, 108}

\bibitem[\protect\citeauthoryear{{Shaw} et~al.,}{{Shaw} et~al.}{2018}]{sls18}
{Shaw} B.,  et~al., 2018, \mn@doi [\mnras] {10.1093/mnras/sty1294}, \href
  {http://adsabs.harvard.edu/abs/2018MNRAS.478.3832S} {478, 3832}

\bibitem[\protect\citeauthoryear{{Shaw} et~al.,}{{Shaw} et~al.}{2022}]{shaw22}
{Shaw} B.,  et~al., 2022, \mn@doi [\mnras] {10.1093/mnras/stac1156}, \href
  {https://ui.adsabs.harvard.edu/abs/2022MNRAS.tmp.1132S} {arXiv:2204.10767v2}

\bibitem[\protect\citeauthoryear{{Shemar} \& {Lyne}}{{Shemar} \&
  {Lyne}}{1996}]{sl96}
{Shemar} S.~L.,  {Lyne} A.~G.,  1996, \mnras, \href
  {http://adsabs.harvard.edu/abs/1996MNRAS.282..677S} {282, 677}

\bibitem[\protect\citeauthoryear{{Sourie} \& {Chamel}}{{Sourie} \&
  {Chamel}}{2020}]{sourie20}
{Sourie} A.,  {Chamel} N.,  2020, \mn@doi [\mnras] {10.1093/mnrasl/slaa015},
  \href {https://ui.adsabs.harvard.edu/abs/2020MNRAS.493L..98S} {493, L98}

\bibitem[\protect\citeauthoryear{{Takata} et~al.,}{{Takata}
  et~al.}{2020}]{twl20}
{Takata} J.,  et~al., 2020, \mn@doi [\apj] {10.3847/1538-4357/ab67b1}, \href
  {https://ui.adsabs.harvard.edu/abs/2020ApJ...890...16T} {890, 16}

\bibitem[\protect\citeauthoryear{{Warszawski}, {Melatos}  \&
  {Berloff}}{{Warszawski} et~al.}{2012}]{warszawski12}
{Warszawski} L.,  {Melatos} A.,   {Berloff} N.~G.,  2012, \mn@doi [\prb]
  {10.1103/PhysRevB.85.104503}, \href
  {https://ui.adsabs.harvard.edu/abs/2012PhRvB..85j4503W} {85, 104503}

\bibitem[\protect\citeauthoryear{{Weltevrede}, {Johnston}  \&
  {Espinoza}}{{Weltevrede} et~al.}{2011}]{wje11}
{Weltevrede} P.,  {Johnston} S.,   {Espinoza} C.~M.,  2011, \mn@doi [\mnras]
  {10.1111/j.1365-2966.2010.17821.x}, \href
  {https://ui.adsabs.harvard.edu/abs/2011MNRAS.411.1917W} {411, 1917}

\bibitem[\protect\citeauthoryear{{Yakovlev} \& {Pethick}}{{Yakovlev} \&
  {Pethick}}{2004}]{yakovlev04}
{Yakovlev} D.~G.,  {Pethick} C.~J.,  2004, \mn@doi [\araa]
  {10.1146/annurev.astro.42.053102.134013}, \href
  {https://ui.adsabs.harvard.edu/abs/2004ARA&A..42..169Y} {42, 169}

\bibitem[\protect\citeauthoryear{{Yi} \& {Zhang}}{{Yi} \& {Zhang}}{2015}]{yz15}
{Yi} S.-X.,  {Zhang} S.-N.,  2015, \mn@doi [\mnras] {10.1093/mnras/stv2261},
  \href {https://ui.adsabs.harvard.edu/abs/2015MNRAS.454.3674Y} {454, 3674}

\bibitem[\protect\citeauthoryear{{Yu} \& {Huang}}{{Yu} \&
  {Huang}}{2016}]{huang16}
{Yu} Y.-B.,  {Huang} Y.-F.,  2016, \mn@doi [RAA] {10.1088/1674-4527/16/5/075},
  \href {https://ui.adsabs.harvard.edu/abs/2016RAA....16...75Y} {16, 75}

\bibitem[\protect\citeauthoryear{{Yu} et~al.,}{{Yu} et~al.}{2013}]{yu13}
{Yu} M.,  et~al., 2013, \mn@doi [\mnras] {10.1093/mnras/sts366}, \href
  {https://ui.adsabs.harvard.edu/abs/2013MNRAS.429..688Y} {429, 688}

\bibitem[\protect\citeauthoryear{{Yuan}, {Wang}, {Manchester}  \& {Liu}}{{Yuan}
  et~al.}{2010a}]{ywm10}
{Yuan} J.~P.,  {Wang} N.,  {Manchester} R.~N.,   {Liu} Z.~Y.,  2010a, \mnras,
  \href {http://adsabs.harvard.edu/abs/2010MNRAS.404..289Y} {404, 289}

\bibitem[\protect\citeauthoryear{{Yuan}, {Manchester}, {Wang}, {Zhou}, {Liu}
  \& {Gao}}{{Yuan} et~al.}{2010b}]{yuan10}
{Yuan} J.~P.,  {Manchester} R.~N.,  {Wang} N.,  {Zhou} X.,  {Liu} Z.~Y.,
  {Gao} Z.~F.,  2010b, \mn@doi [\apjl] {10.1088/2041-8205/719/2/L111}, \href
  {https://ui.adsabs.harvard.edu/abs/2010ApJ...719L.111Y} {719, L111}

\bibitem[\protect\citeauthoryear{{Yuan}, {Manchester}, {Wang}, {Wang}, {Zhou},
  {Yan}  \& {Liu}}{{Yuan} et~al.}{2017}]{ymw17}
{Yuan} J.~P.,  {Manchester} R.~N.,  {Wang} N.,  {Wang} J.~B.,  {Zhou} X.,
  {Yan} W.~M.,   {Liu} Z.~Y.,  2017, \mn@doi [\mnras] {10.1093/mnras/stw3203},
  \href {https://ui.adsabs.harvard.edu/abs/2017MNRAS.466.1234Y} {466, 1234}

\bibitem[\protect\citeauthoryear{{Zhao} et~al.,}{{Zhao} et~al.}{2017}]{znl17}
{Zhao} J.,  et~al., 2017, \mn@doi [\apj] {10.3847/1538-4357/aa74d8}, \href
  {https://ui.adsabs.harvard.edu/abs/2017ApJ...842...53Z} {842, 53}

\bibitem[\protect\citeauthoryear{{Zhou} et~al.,}{{Zhou} et~al.}{2019}]{zzz19}
{Zhou} S.~Q.,  et~al., 2019, \mn@doi [\apss] {10.1007/s10509-019-3660-7}, \href
  {https://ui.adsabs.harvard.edu/abs/2019Ap&SS.364..173Z} {364, 173}

\bibitem[\protect\citeauthoryear{{Zou}, {Wang}, {Wang}, {Manchester}, {Wu}  \&
  {Zhang}}{{Zou} et~al.}{2004}]{zww04}
{Zou} W.~Z.,  {Wang} N.,  {Wang} H.~X.,  {Manchester} R.~N.,  {Wu} X.~J.,
  {Zhang} J.,  2004, \mnras, \href
  {http://adsabs.harvard.edu/abs/2004MNRAS.354..811Z} {354, 811}

\makeatother
\end{thebibliography}

\end{document}